# Nonlinear connections and 1+3 splittings of spacetime

**D H Delphenich**
Physics Department, Bethany College, Lindsborg, KS 67456

E-mail: david_delphenich@yahoo.com

**Abstract**. The analogy between 1+3 splittings of the spacetime tangent bundle and the splitting of the tangent bundle to the bundle of linear frames into vertical and horizontal sub-bundles is described from the unifying standpoint of the geometry of foliations. The physical nature of the line field on spacetime that plays the role of vertical sub-bundle is discussed in some detail. The notion that the complementary spatial bundle is most fundamentally a representation of the normal bundle to the foliation of spacetime by the integral curves of the line field is proposed, such that the geometry of space becomes the transverse geometry of the flow and the integration of the spatial bundle into proper time simultaneity hypersurfaces becomes a secondary issue to the geometry of space as the geometry of the leaf space of the foliation. The concept of an adapted convected frame field is introduced as a means of locally representing the transversal geometry of a flow in terms of the information that is contained in the flow and its derivatives, and some discussion is given to the role of the Bott connection.

## Contents

## 1. Introduction.

One of the fundamental dilemmas of all science is to transcend the subjectivity of empirical observations. In particular, one of the key contributions of the theory of relativity to physics was the idea that motion is better described by a four-dimensional manifold than a three-dimensional one. However, experimental physics is still limited by the fact that all measurements are made in the rest space of the measuring devices. Hence, one can directly measure only the proper time parameter associated with the motion of the measuring devices and not the fourth dimension of spacetime, more generally.

    For that matter, even theoretical physics is limited by the fact that the human brain perceives motion by way of its projection into the three-dimensional space of one's spatial intuition. In fact, one's vision projects further into a projective plane and infers the third dimension by way of perspective, precedence, and parallax, so the extension beyond the dimensions of perception must be made by abstraction, generalization, and idealization. Here, one must accept that many possible avenues of generalization must exist that all project back into one's perceptions the same way.

    The approach taken in this article is to start with the four-dimensional picture and examine the manner by which one deduces a three-dimensional picture by essentially "modding out" the motion of the observer/measurer. We shall start by assuming that a physical motion without fixed points is given in the form of a foliation of dimension one in a four-dimensional manifold $M$, and that the three-dimensional picture is obtained only locally by projection in the tangent spaces onto transversal subspaces. The existence of a three-dimensional spatial manifold would follow only by the integrability of the rank-three sub-bundle of the tangent bundle onto which we are projecting. The spatial manifold would then take the form of a leaf of a foliation that is transverse to the motion. Even this sort of construction leaves open the possibility that the topology of the leaf might change along the motion.

    An obvious issue to address in this construction is the fact that transversal subspaces are not going to be unique, in general. Certainly, if one assumes that $M$ is Lorentzian, as is customary in relativity theory, one can single out a unique orthogonal complement to the line field $L(M)$ that is tangent to the motion, but if one wishes to remain faithful to the spirit of pre-metric physics, as is suggested by the study of pre-metric electromagnetism, one might wish to see how much can be said up to that point. In particular, one might start with only a Riemannian structure on the normal bundle to the motion and look at how one extends this to the Lorentzian structure on $M$ by appropriate physical axioms. Hence, we shall regard a spatial tangent sub-bundle $\Sigma(M)$ in $T(M)$ that is complementary to $L(M)$ as merely a convenient representation of the normal bundle to the foliation.

    We shall not assume that the spatial bundle is integrable, because that depends upon the nature of the motion, and, in the Lorentzian case, the nature of the metric, as well.

We shall regard such a simplification as a limiting approximation that is generally possible only locally, but generally obstructed by geometry and topology. This is somewhat equivalent to assuming that "inertial reference frames" are also physical approximations, just as static fields are. This is based in the idea that in the absence of an absolute rest frame for motion in the universe all reference frames are inherently moving, if not accelerated to some degree. Whether or not the acceleration plays a significant role in the physical measurements is not as important as whether the motion defines spatial frames that are ultimately integrable, even though the obstruction to integrability might have to cosmological in character. For instance, one notes that rotational motions seem unavoidable in the astronomical context, from the rotation of the Earth about its axis, to its orbit around the Sun, to the orbiting of the stars about the galactic centre, and so on.

Of course, from the standpoint of experiment the issue is one of how long the experiment is to last. Over a time span of seconds, even the rotation of the Earth may be negligible, although one notices that over a time span of minutes this motion will affect the observations one makes through a telescope that is not synchronized with the Earth rotation, and over the span of a day, one will have to contend with non-negligible Foucault precession. However, in the eyes of topology, which is insensitive to matters of scale, the issue of time scale is irrelevant compared to the issue of whether singularities exist.

An interesting physical subtlety that is associated with regarding all motions as potentially non-integrable is that it suggests that the concept of a local frame field is more fundamental to physics than that of a local coordinate system. This is, of course, completely contrary to the historical path that was followed in the relativistic treatment of spacetime structure, which always started with coordinate systems as the fundamental physical constructions. What we are suggesting is that coordinate systems are obtained by the integration of holonomic local frame fields, and that the assumption of holonomicity is an approximation that is equivalent to the assumption of vanishing acceleration.

The basic subject of this article is related to the treatment of the transversal geometry of the one-dimensional foliation that is defined by a physical motion as a generalization of the geometry of linear connections on a bundle of linear frames to nonlinear connections on a manifold that is given an almost-product structure that has the same integrability properties as the vertical and horizontal splitting of the tangent bundle to a frame bundle; viz., the vertical sub-bundle is integrable, but not necessarily the horizontal one. We shall call such an almost-product structure *semi-holonomic;* a local frame field that is adapted to such an almost-product structure will also be necessarily semi-holonomic.

Some of the methodology of the geometry of almost-product structures was previously developed by Cattaneo [1], who defined transverse and longitudinal derivatives, both ordinary and covariant, and applied them to some decompositions of interest to general relativity, such as the vorticity tensor and the Killing tensor for a vector field. Much of it goes back to work of Vranceanu [2] on non-holonomic spaces. Some of the methodology has also been applied to the specific problem of expressing the 1+3 form of the Maxwell equations in an anholonomic frame field, as in Kocik [3] and Fecko [4]. The terminology of nonlinear connections is largely borrowed from Vacaru and others [5]. However, the approach of using the transversal geometry of foliations is original to this work, to the author's knowledge, and, in particular, the idea that the three-dimensional space of one's intuition and observations represents the leaf space of a one-dimensional foliation that gets defined by the motion of the observer. Some discussion is also given here to the role of the Bott connection for the foliation, which is usually only mentioned in the context of characteristic classes for foliations.

Since the basic object on which the rest of the analysis is founded is a one-dimensional foliation that represents the motion of an observer, a key objective of this approach is to make the transversal geometry reflect, to the greatest degree possible, the information that is contained in that foliation. Hence, considerable attention is paid in this work to the sort of information that such a foliation might possess and how it gets represented into the transversal tangent spaces. In particular, the notion of an adapted convected frame field is introduced as a means of spanning the transversal subspaces with a frame field that is defined by the local flow of the vector field that generates the foliation. The torsion and curvature of the transversal bundle are then locally expressible in terms of the derivatives of that flow.

The basic structure of this article is then to first summarize the key notions concerning the integrability of sub-bundles and foliations in section 2. We then discuss semi-holonomic frame fields as a convenient way of addressing the local constructions in section 3, and use the example of a linear connection in section 4 as a way of illustrating the geometry of a semi-holonomic almost-product structure in an established context. After defining almost-product structures in section 5 and discussing some of the associated ideas, the concept of a nonlinear connection as a generalization of the aforementioned linear connections is detailed in section 6. The physical issues concerning physical observers and rest spaces are presented in section 7 in a manner that is specifically directed to the nature of almost-product structures, and then the general geometry of nonlinear connections is presented in section 8. Finally, the foregoing constructions and results are specialized to the case of the spacetime manifold when it is given a one-dimensional foliation in section 9. Further avenues for exploration and development are proposed in section 10.

## 2. The integrability of differential systems.

A *differential system* on a differentiable manifold $M$ is a sub-bundle $\mathcal{D}(M)$ of the tangent bundle $T(M)$ that has constant rank. That is, to every point $x \in M$ one associates a vector subspace $\mathcal{D}_x(M)$ of $T_x(M)$, which one refers to as an *integral element*, and all of these subspaces have the same dimension. The term "distribution" on $M$ is often used, but we shall not use it here, since we shall also consider distributions of mass in the other sense of the term "distribution."

A differential system $\mathcal{D}(M)$ is called *integrable* iff $M$ can be foliated by integral submanifolds of $\mathcal{D}(M)$. A submanifold $i:L \to M$ is an *integral submanifold* of $\mathcal{D}(M)$ iff the tangent space $i_*T_u(L)$ is a vector subspace of $\mathcal{D}_{i(u)}(M)$ for every $u \in L$. The differential system $\mathcal{D}(M)$ is called *completely integrable* iff there is an integral submanifold of maximal dimension through each point of $M$. Hence, for the completely integrable case, the dimension of any integral submanifold is equal to the dimension of the fibres of $\mathcal{D}(M)$. Conceivably, a differential system that is not completely integrable might still admit integral submanifolds of lower dimension than the rank of $\mathcal{D}(M)$.

To be more precise about the term "foliation" we clarify that a *foliation* of dimension *n* of an *m+n*-dimensional manifold *M* is a decomposition of *M* into a disjoint union of subsets – called the *leaves* of the foliation – that are each submanifolds of *M* of dimension *n*. Furthermore, every point of *M* has a coordinate chart [1] $(U, x^i, x^a)$ that is *adapted* to this foliation in that the local piece of any leaf through *U* is described by the equations $x^i =$ constant [2]. Since this is always possible, we further demand that when two such coordinates $(U, x^i, x^a)$ and $(V, \bar{x}^j, \bar{x}^b)$ overlap, in the intersection $U \cap V$ the coordinate transformation must take adapted coordinates to adapted coordinates; i.e., it must preserve the leaves. This condition can be expressed as the system of equations:

$$\bar{x}^j = \bar{x}^j(x^i, x^a), \qquad \bar{x}^b = \bar{x}^b(x^a). \qquad (2.1)$$

Hence, the differential of this transformation has the matrix form:

$$\frac{\partial \bar{x}^I}{\partial x^J} = \begin{bmatrix} \dfrac{\partial \bar{x}^j}{\partial x^i} & \dfrac{\partial \bar{x}^j}{\partial x^a} \\ 0 & \dfrac{\partial \bar{x}^b}{\partial x^a} \end{bmatrix}. \qquad (2.2)$$

Since this matrix must be invertible at all points of $U \cap V$ one sees that it must take its values in the subgroup $GL(n+m; n)$ of $GL(n+m)$ that preserves the $\mathbb{R}^n$ subspace of vectors of the form $(0, \ldots, 0, v^{m+1}, \ldots, v^{m+n})$.

One can also define local frame fields and coframe fields on *M* that are adapted to the foliation – or, more generally, to a differential system $\mathcal{D}(M)$ – by saying that an *m+n* frame field $\{\mathbf{e}_i, \mathbf{e}_a\}$ on $U \subset M$ is *adapted* to the differential system $\mathcal{D}(M)$ iff the *n*-frame field $\mathbf{e}_a$ spans the vector space $\mathcal{D}_x(M)$ at every point $x \in U$; hence, it is tangent to the leaves at every point. For instance, the natural coordinate frame field $\{\partial_i, \partial_a\}$ for an adapted coordinate chart will be an adapted local frame field. When the domains *U* and *V* of two adapted local frame fields overlap the transition function must take its values in $GL(n+m; m)$.

One can also make statements that are dual to the preceding ones in terms of local coframe fields. A local *m+n*-coframe field $\{\theta^i, \theta^a\}$ on $U \subset M$ is *adapted* to a differential system $\mathcal{D}(M)$ iff the local $m$-coframe field $\theta^i$ has $\mathcal{D}_x(M)$ for its annihilating subspace at every $x \in U$. This says that locally we can locally represent the differential system $\mathcal{D}(M)$ by an *exterior differential system* of the form:

$$\theta^i = 0. \tag{2.3}$$

---

[1] We shall use the convention that if *M* is *m+n*-dimensional then the first coordinate indices will be letters *i*, *j*, *k*, … that range from 1 to *m* and the second coordinate indices will be letters *a*, *b*, *c*, … that range from *m*+1 to *n*. Upper case Latin indices range from 1 to *n+m*. Lower case Greek letters will range from 0 to *m+n*−1.

[2] This convention is slightly non-standard, but we shall be considering foliations in which it is more convenient to regard the second set of coordinates as parameterizing the leaves.

Since the exterior forms in this system are all 1-forms, one often hears it referred to as a *Pfaffian* system.

We point out the computational subtlety that this system of equations is actually a system of *algebraic* equations, not a system of differential equations. The solution to this system is a sub-bundle of $T(M)$ whose fibres are the integral elements to a differential system $\mathcal{D}(M)$. If $\{\mathbf{E}_I, I = 1, \ldots, m+n\}$ is an arbitrary $m+n$-frame field on $U \subset M$ and $\{E^I, I = 1, \ldots, m+n\}$ is its *reciprocal* coframe field – i.e., $E^I(\mathbf{E}_J) = \delta^I_J$ – then the 1-forms $\theta^i$ can be written as:

$$\theta^i = F^i_I E^I. \tag{2.4}$$

A tangent vector $\mathbf{X} = X^I \mathbf{E}_I$ is a solution of (2.3) iff $i_\mathbf{X} \theta^i = 0$ for all $i$, which means its components $X^I$ are solutions of the system of linear equations:

$$F^i_I X^I = 0, \qquad \text{for all } i. \tag{2.5}$$

By assumption, this system will have $n$ linearly independent solutions, whose $n \times (n+m)$ matrix of components we shall call $A^I_a$.

The system of partial differential equations then appears locally in the form of the equations for an integral submanifold $\iota: \mathbb{R}^n \to U \subset M$, $u^a \mapsto Y^I(u^a)$ to $\mathcal{D}(M)$, which has the form:

$$\frac{\partial Y^I}{\partial u^a} = A^I_a. \tag{2.6}$$

One might think that it would be more efficient to use adapted coordinates in these computations from the outset. However, in effect, one must integrate the equations (2.6) to find adapted coordinates to begin with.

It will be essential in what follows for us to have the necessary and sufficient conditions for a differential system $\mathcal{D}(M)$ to be completely integrable. These conditions take the form of *Frobenius's theorem*, which can be expressed in various forms. We shall confine ourselves to the forms that we shall use in what follows.

In terms of $\mathcal{D}(M)$ itself, if we let $\mathfrak{X}(\mathcal{D})$ represent the (infinite-dimensional) vector space of sections of $\mathcal{D}(M)$ – i.e., vector fields on $M$ that take their values in the fibres of $\mathcal{D}(M)$ – then Frobenius's theorem says that $\mathcal{D}(M)$ is completely integrable iff $\mathfrak{X}(\mathcal{D})$ is also a Lie sub-algebra of $\mathfrak{X}(M)$ under the Lie bracket of vector fields; this property is generally referred to as *involutivity*. Consequently, if $\{\mathbf{e}_a, a = 1, \ldots, n\}$ is a local $n$-frame field on $U \subset M$ that consists of vector fields in $\mathfrak{X}(\mathcal{D})$ then there must exist a set of functions $c_{ab}^c$, $a, b, c = 1, \ldots, n$ on $U$ such that:

$$[\mathbf{e}_a, \mathbf{e}_b] = c_{ab}^c \mathbf{e}_c. \tag{2.7}$$

One calls these functions the *structure functions* of the local frame field $\mathbf{e}_a$.

If the differential system $\mathcal{D}(M)$ is defined by an exterior differential system of the Pfaffian form (2.3) on $U$ then the dual statement to (2.7) is that $\mathcal{D}(M)$ is completely integrable iff there exist functions $c_{jk}^i$ on $U$ such that:

$$d\theta^i = -\tfrac{1}{2} c_{jk}^i \theta^j \wedge \theta^k. \tag{2.8}$$

The functions $c_{jk}^i$ are then the structure functions of the coframe field $\theta^i$; in general, they will differ from the $c_{ab}^c$. However, if $\theta^a$ is the reciprocal coframe field to $\mathbf{e}_a$ then by applying the intrinsic definition of the exterior derivative formula (see Sternberg [6]):

$$d\theta^c(\mathbf{e}_a, \mathbf{e}_b) = \mathbf{e}_a(\theta^c(\mathbf{e}_b)) - \mathbf{e}_b(\theta^c(\mathbf{e}_a)) - \theta^c[\mathbf{e}_a, \mathbf{e}_b] = -\theta^c[\mathbf{e}_a, \mathbf{e}_b], \tag{2.9}$$

one finds that (2.7) is equivalent to:

$$d\theta^c = -\tfrac{1}{2} c_{ab}^c \theta^a \wedge \theta^b. \tag{2.10}$$

The difference between (2.8) and (2.10) is the difference between annihilating an $n$-dimensional tangent subspace and spanning an $n$-dimensional cotangent subspace, respectively.

It is no coincidence that both equations (2.8) and (2.10) look similar in form to the Maurer-Cartan equations of the theory of Lie groups. One must realize that the integral submanifolds of a rank $n$ differential system on an $n$-dimensional manifold will be zero-dimensional; i.e., they will be the points of the manifold. In such a case, the question of integrability reduces to a triviality and the equations become a property of any frame field. Indeed, the equations in question are also valid for a parallelizable manifold that does not have to possess a group structure. The difference is that there can be a global frame field that has constant structure functions on a parallelizable manifold iff the manifold also has a group structure, at least locally.

From the preceding remarks, it becomes clear that the integrability of a sub-bundle of a tangent bundle becomes an issue only when it has less than maximal rank. In the works of Vranceanu [2], Horak [7], Synge [8], Schouten [9], and others, what we are calling a non-integrable differential system was referred to as a *non-holonomic* (or *anholonomic*) *space*. This derives from the fact that one refers to a local frame field $\mathbf{e}_i$ as *holonomic* iff:

$$[\mathbf{e}_i, \mathbf{e}_j] = 0 \tag{2.11}$$

and *anholonomic,* otherwise. Dually, a local coframe field $\theta^i$ is holonomic iff:

$$d\theta^i = 0 \tag{2.12}$$

and anholonomic, otherwise. If $\theta^i$ is the reciprocal frame field to $\mathbf{e}_i$, so – $\theta^i(\mathbf{e}_j) = \delta^i_j$ – then (2.12) follows from the intrinsic formula (2.9) for $d\theta^i$ and the fact that $\theta^k$ is a linear isomorphism of tangent spaces to $\mathbb{R}^n$.

This present terminology is consistent with the usage of the terms "holonomic" and "anholonomic" in physics since it explains both the concept of (an)holonomic constraints in mechanics and (an)holonomic frame fields in relativity. In the former case, if a configuration space for a physical system is described by a manifold then holonomic constraints on this configuration manifold define a sub-bundle of the tangent bundle that will integrate into a constraint submanifold.

A holonomic frame field in relativity is what one usually hears referred to as an "inertial" frame field or inertial coordinate system. An anholonomic frame field might then take the form of a "non-inertial coordinate system," which is really a misnomer, since one does not have the integrability that would allow one to define a coordinate system whose natural frame field $\partial_i$ would agree with the given frame field; i.e., all natural frame fields are holonomic.

One of the key features of a foliation on a manifold $M$ is that it distinguishes only one sub-bundle of $T(M)$, namely, the sub-bundle $\mathcal{D}(M)$, but does not necessarily give any indication of how one might define a complementary sub-bundle. Instead, one deals with the *normal bundle* to $\mathcal{D}(M)$ as a sort of complement. The normal bundle $Q(M)$ to $\mathcal{D}(M)$ is defined to be the quotient bundle whose fibre $Q_x(M)$ at each $x \in M$ is equal to $T_x(M)/\mathcal{D}_x(M)$; i.e., the translates of the subspace $\mathcal{D}_x(M)$ by the other vectors in $T_x(M)$. If $M$ is $m+n$-dimensional and the dimension of the leaves of the foliation is $n$ then the fibres of $Q(M)$ will be vector spaces of dimension $m$. One can also regard the elements of each $Q_x(M)$ as equivalence classes of tangent vectors $\mathbf{v}, \mathbf{w} \in T_x(M)$ under the equivalence $\mathbf{v} \sim \mathbf{w}$ iff $\mathbf{v} - \mathbf{w} \in \mathcal{D}_x(M)$. This defines a canonical projection $q: T(M) \to Q(M)$ that takes each tangent vector to its equivalence class.

There is a linear isomorphism between each $Q_x(M)$ and the sub-bundle of $T_x^*(M)$ that consists of all covectors that annihilate the vector spaces $\mathcal{D}_x(M)$. It is obtained from the fact that if $\mathbf{v} - \mathbf{w} \in \mathcal{D}_x(M)$ and $\alpha \in T_x^*(M)$ such that $\alpha$ annihilates the vectors in $\mathcal{D}_x(M)$ then $\alpha(\mathbf{v}) = \alpha(\mathbf{w}) = a_0$, which is the same for all vectors in the equivalence class $[\mathbf{v}]$ of $\mathbf{v}$. Moreover, there is only one $\alpha$ with this property since we have specified the value of $\alpha$ on all vectors in $T_x(M)$. Hence, the isomorphism takes $[\mathbf{v}]$ to that covector $\alpha \in T_x^*(M)$ such that $\alpha(\mathbf{v}) = a_0$ and $\alpha(\mathbf{x}) = $ when $\mathbf{x} \in \mathcal{D}_x(M)$.

It is also common to refer to the sub-bundle $\mathcal{D}(M)$ as the *structural* bundle and the bundle $Q(M)$ as the *transversal* one.

The dominant approach to the geometry of foliations is to examine how the geometrical and topological information in the bundle $\mathcal{D}(M)$ gets represented in its normal bundle; one calls the geometry of the normal bundle *transversal geometry*. For instance, a foliation is called *Riemannian* if one has defined a Riemannian structure on its normal bundle. In effect, one avoids the necessity of dealing with the *leaf space* of the foliation, which is the quotient of the manifold by the equivalence relation of membership in the same leaf and can be a topologically troublesome space; in particular, it does not have to be a manifold. This is analogous to the way that the theory of transformation groups avoids the necessity of dealing with the equally pathological orbit space of a group action by considering the transversal geometry to the orbits. In the case of a Lie foliation – viz., one whose leaves are the orbits of a Lie group action – these two notions are identical. For instance the fibres of a principle fibre bundle can also be regarded as leaves of a foliation of the total space, as well as orbits of the action of the structure group.

### 3. Semi-holonomic frame fields.

Just as the diagonal form for a diagonalizable matrix contains the minimum amount of information that it takes to describe it, one often wishes to find local frame fields that are as natural to a given geometric structure as possible. In this section, we shall define and discuss a certain type of local frame field that one uses when the geometrical structure is a foliation of dimension $n$ in an $m+n$-dimensional manifold $M$.

The most natural sort of adapted local frame field on a foliated manifold is a frame field on $U \subset M$ that reflects the integrability of the sub-bundle $\mathcal{D}(M)$, i.e., the tangent bundle to the leaves. A *semi-holonomic frame field* on $U \subset M$ is an adapted frame field $\{\mathbf{e}_i, \mathbf{e}_a\}$ such that $\mathbf{e}_i$ is anholonomic and $\mathbf{e}_a$ is holonomic. One cannot hope to find an adapted local coordinate system such that both $\mathbf{e}_i$ and $\mathbf{e}_a$ are holonomic unless $M$ has a transversal foliation along with the one that we are assuming. Our immediate problem is to find the simplest sort of frame transformation that will take the natural frame field of an adapted coordinate system $(U, x^i, x^a)$ to a semi-holonomic frame field on $U$.

Suppose we have an adapted local coordinate chart $(U, x^i, x^a)$ such that the $n$-frame $\partial_a$ spans $\mathcal{D}(U)$, while the $m$-frame $\partial_i$ spans some other transversal complement to $\mathcal{D}(U)$, which we denote by $\mathcal{D}^c(U)$. Consequently, both of the local frame fields $\partial_i$ and $\partial_a$ are holonomic.

Now, suppose we have a non-integrable complement $H(M)$ to $\mathcal{D}(M)$. If $(U, x^i, x^a)$ is an adapted coordinate chart for this splitting of $T(M)$ and $\{\mathbf{e}_i, \mathbf{e}_a\}$ is a semi-holonomic frame field on $U$ then the simplest form that the transformation from the natural frame to the semi-holonomic frame can take amounts to projecting the basis vectors in $\mathcal{D}^c(U)$ onto vectors in $H(M)$. Since any tangent vector $\mathbf{v} = \mathbf{v}_H + \mathbf{v}_V$, if the projection $N: \mathcal{D}^c(U) \to V(U)$ has a matrix relative to the natural frame that we denote by $N_i^a$ then the simplest form that such projection may take is:

$$\mathbf{e}_i = \partial_i - N_i^a \partial_a, \qquad \mathbf{e}_a = \partial_a. \qquad (3.1)$$

We easily obtain the structure functions of a semi-holonomic frame from (3.1):

$$[\mathbf{e}_i, \mathbf{e}_j] = -(\mathbf{e}_i N_j^a - \mathbf{e}_j N_i^a)\mathbf{e}_a, \quad [\mathbf{e}_a, \mathbf{e}_i] = -(\mathbf{e}_a N_j^b)\mathbf{e}_b, \quad [\mathbf{e}_a, \mathbf{e}_i] = 0. \tag{3.2}$$

This makes the structure functions equal to:

$$c_{ij}^a = -(\mathbf{e}_i N_j^a - \mathbf{e}_j N_i^a), \qquad c_{ai}^b = -\mathbf{e}_a N_j^b, \qquad c_{ij}^k = c_{bc}^a = c_{bc}^i = c_{ai}^j = 0. \tag{3.3}$$

The reciprocal situation in terms of the coframe field $\{dx^i, dx^a\}$ is that $dx^a$ cannot annihilate $H^*(U)$, since it annihilates $\mathcal{D}^c(U)$. The local coframe field $\{\theta^i, \theta^a\}$ that is reciprocal to $\{\mathbf{e}_i, \mathbf{e}_a\}$ then takes the form:

$$\theta^i = dx^i, \qquad \theta^a = dx^a + N_i^a dx^i. \tag{3.4}$$

This has the immediate consequence that we can identify the matrix $N_i^a$ as representing:

$$N_i^a = \theta^a(\partial_i). \tag{3.5}$$

That is, if we define $H(U)$ by the annihilating subspaces of the 1-forms $\theta^a$ then the matrix $N_i^a$ is also defined by their components relative to $dx^i$. The vertical components can be normalized to one or zero since the vertical parts are necessarily non-zero and the equations $\theta^a(\mathbf{v}) = 0$ are homogeneous.

From (3.3), this makes the structure equations for $\{\theta^i, \theta^a\}$ take the form:

$$d\theta^i = 0, \qquad d\theta^a = -\tfrac{1}{2} c_{ij}^a \theta^i \wedge \theta^j - \tfrac{1}{2} c_{ib}^a \theta^i \wedge \theta^b. \tag{3.6}$$

Clearly, in such a frame field all of the non-trivial information about $H(U)$ is contained in $d\theta^a$.

Let us look closer at the nature of the transformation (3.1) from a natural frame to a semi-holonomic one. The transition function $A: U \to GL(n)$ that it is associated with it takes its values in matrices of the form:

$$[A]_J^I = \left[\begin{array}{c|c} \delta_j^i & -N_j^a \\ \hline 0 & \delta_b^a \end{array}\right]. \tag{3.7}$$

Since the product of two such matrices $A$ and $B$ is again of the same form:

$$[AB]_J^I = \left[\begin{array}{c|c} \delta_j^i & A_j^a + B_j^a \\ \hline 0 & \delta_b^a \end{array}\right] \tag{3.8}$$

and $\det(A) = 1$, we see that the set of all such matrices defines a subgroup of $SL(n)$ of dimension $mn$ that is isomorphic to $\mathbb{R}^{mn}$; hence, it is Abelian.

Of particular interest is what happens to the components of the submatrix $N_j^a$ under a change of adapted coordinate chart. Such a coordinate transformation takes the form (2.1), so the semi-holonomic frame $\{\mathbf{e}_i, \mathbf{e}_a\}$ then transforms to the new semi-holonomic frame $\{\overline{\mathbf{e}}_i, \overline{\mathbf{e}}_a\}$:

$$\overline{\mathbf{e}}_i = \frac{\partial \overline{x}^j}{\partial x^i} \mathbf{e}_j = \overline{\partial}_i - \overline{N}_i^a \overline{\partial}_a, \tag{3.9}$$

in which:

$$\overline{N}_i^a = \frac{\partial x^j}{\partial \overline{x}^i} N_j^b \frac{\partial \overline{x}^a}{\partial x^b} + \frac{\partial x^j}{\partial \overline{x}^i} \frac{\partial \overline{x}^a}{\partial x^j}, \tag{3.10}$$

which has an inhomogeneous character that is closely analogous to the transformation law for the components of a connection 1-form in a natural frame.

**4. Linear connections.**
There are many ways of characterizing a connection on a fibre bundle, and the choice is usually dictated by the nature of the problem that one is addressing. Now, the nature of the problem at hand is defined by the assumption that the four-dimensional spacetime manifold $M$ is endowed with a Whitney sum splitting $L(M) \oplus \Sigma(M)$ of its tangent bundle $T(M)$ into and a real line bundle $L(M)$ whose one-dimensional fibres are identified as "temporal" and a rank-three sub-bundle $\Sigma(M)$ whose fibres are identified as "spatial." Hence, the approach to connections and their generalizations that we shall take is that of Whitney sum splittings of tangent bundles into vertical and horizontal sub-bundles. We will then treat the spatial sub-bundle as the analogue of the horizontal one and the temporal sub-bundle as the analogue of the vertical one.

Furthermore, we shall start with the particular case of a fibre bundle that takes form of the $GL(n)$-principal bundle $GL(M)$ of linear frames on an $n$-dimensional manifold $M$. The fibres of this bundle are then all diffeomorphic to $GL(n)$ as manifolds, but not canonically. One can also regard them as the orbits of the right action of $GL(n)$ on the manifold $GL(M)$, or as the leaves of a foliation, which we shall now discuss.

The bundle $GL(M)$ has some canonical constructions associated with it that will need to be generalized. The first one is a sub-bundle of the tangent bundle $T(GL(M))$ to the total space that one calls the *vertical* sub-bundle $V(GL(M))$, and which we abbreviate by $V$. It is defined by the kernel of the map $\pi_*: T(GL(M)) \to T(M)$, which then consists of all vectors in $T(GL(M))$ that are tangent to the fibres. The bundle $V$ has the property that it is integrable when regarded as a differential system on $GL(M)$, and the integral submanifolds are the fibres of $GL(M)$. This is due to the fact that the vector space $\mathfrak{X}(V)$ of vertical vector fields on $GL(M)$ is a Lie sub-algebra of the Lie algebra $\mathfrak{X}(GL(M))$ of all vector fields on $GL(M)$.

Hence, since $V$ is also trivializable – as is $T(GL(M))$ itself – if we define a trivializing $n^2$-frame field $\{\mathbf{V}_I, I = 1, \ldots, n^2\}$ for $V$ then there are smooth functions $c_{IJ}^K$ on $GL(M)$ such that:

$$[\mathbf{V}_I, \mathbf{V}_J] = c_{IJ}^K \mathbf{V}_K. \tag{4.1}$$

We also point out that the right action of *GL*(*n*) defines fundamental vector fields on *GL*(*M*) that are all vertical and define a representation of the Lie algebra $\mathfrak{gl}(n)$ in $\mathfrak{X}(V)$. If the trivializing frame field **V**$_I$ consists of the fundamental vector fields that correspond to a basis for $\mathfrak{gl}(n)$ then the structure functions $c_{IJ}^K$ are the structure constants of the Lie algebra $\mathfrak{gl}(n)$ for this basis. Hence, there is a faithful representation of $\mathfrak{gl}(n)$ in $\mathfrak{X}(V)$.

The second canonical construction on *GL*(*M*) is an $\mathbb{R}^n$-valued 1-form $\theta^\mu$. It originates in the fact that a linear frame in $T_x(M)$ can be defined as a linear isomorphism **e**$_x$: $\mathbb{R}^n \to T_x(M)$, $v^\mu \mapsto v^\mu \mathbf{e}_\mu$. The inverse isomorphism $\theta^\mu$: $T_x(M) \to \mathbb{R}^n$, $\mathbf{v} \mapsto \theta^\mu(\mathbf{v}) = v^\mu$ then defines the reciprocal coframe $\theta^\mu$; hence $\theta^\mu(\mathbf{e}_\nu) = \delta^\mu{}_\nu$. One can then regard a coframe on $T_x(M)$ as a 1-form with values in $\mathbb{R}^n$. If one pulls this 1-form in $T_x(M)$ up to $GL_\mathbf{e}(M)$ by means of the projection $GL(M) \to M$ and does so for every frame $\mathbf{e}_x \in GL(M)$ then one eventually defines an $\mathbb{R}^n$-valued 1-form on *GL*(*M*) that we also denote by $\theta^\mu$. It has the property that if $\mathbf{e}_\mu: U \to GL(M)$, $x \mapsto \mathbf{e}_\mu(x)$ is a local frame field then the pull-down of $\theta^\mu$ to *U* is the reciprocal coframe field to $\mathbf{e}_\mu$. Hence, we are not abusing our notation too grievously by using the same symbol for both the canonical 1-form and its local representative on *U*.

The canonical 1-form $\theta^\mu$ is clearly contragredient under the action of *GL*(*n*), so if $h \in GL(n)$ then:

$$\theta^\mu_{\mathbf{e}h} = h^{-1} \theta^\mu_{\mathbf{e}}. \tag{4.2}$$

This amounts to either the statement that a coframe transforms by the inverse of a frame transformation, or that the components of a tangent vector also transform by the inverse of that transformation.

The tangent subspaces to *GL*(*M*) that are annihilated by $\theta^\mu$ are precisely the vertical subspaces. Hence:

$$\theta^\mu(\mathbf{v}) = 0 \qquad \text{iff} \qquad \mathbf{v} \in V. \tag{4.3}$$

The integrability condition for the exterior differential system $\theta^\mu = 0$ on *GL*(*M*) is that there must be a set of $n^2$ $\mathbb{R}^n$-valued 1-forms of the form:

$$\eta^\mu_\nu = \eta^\mu_{\lambda\nu} \theta^\lambda \tag{4.4}$$

such that:

$$d\theta^\mu = \tfrac{1}{2} \eta^\mu_{\lambda\nu} \theta^\lambda \wedge \theta^\nu. \tag{4.5}$$

This integrability condition is dual to the integrability condition (4.1), although in order to relate the two sets of structure functions, we must move on to linear connections.

A *linear connection* on *GL*(*M*) is defined by a Whitney sum splitting $H \oplus V$ of $T(GL(M))$. In this splitting, the complementary *horizontal* sub-bundle *H*, which is short for *H*(*GL*(*M*)), is not usually canonically defined, but must be chosen. Furthermore, it is assumed to have the property that it is invariant under the right action of *GL*(*n*) on *GL*(*M*); that is, if $h \in GL(n)$ takes $\mathbf{e} \in GL(M)$ to $\mathbf{e}h \in GL(M)$ then the differential of the right-translation map takes the fibre $H_\mathbf{e}$ isomorphically to the fibre $H_{\mathbf{e}h}$, as well. Since the dimension of the fibres of *H* will be the same as the dimension of the fibres of *T*(*M*) one can envision the horizontal subspaces as essentially "lifts" of the tangent space to *M* to tangent spaces to *GL*(*M*). The nature of the projection of *H* onto *T*(*M*) and its singularities has much to say about the nature of the connection that one defines, and is obstructed by the topology of *M*. In particular, this projection is trivial only if *M* is parallelizable.

By analogy with the more general case, since the vertical sub-bundle is integrable, we can regard it as the structural bundle of the foliation of *GL*(*M*) by its fibres and the horizontal complement as the transversal one.

As pointed out above, the manifold *GL*(*M*) is always parallelizable. Indeed, this is equivalent to the fact that at least one linear connection always exists on it (see Sternberg [8]). Hence, one can regard *GL*(*M*) as a sort of "parallelizable covering manifold" for *M* in which one "unfolds" the singularities of the frame fields on *M* in the extra dimensions of the fibres.

Since we already discussed the trivialization of *V* above, we shall now address the trivialization of *H*. This would entail the existence of *n* linearly independent horizontal vector fields $\{\mathbf{H}_\mu, \mu = 1, \ldots, n\}$ on *GL*(*M*). Since the canonical 1-form $\theta^\mu$ has the property that it defines a linear isomorphism of each $H_\mathbf{e}$ with $\mathbb{R}^n$ – i.e., a coframe on the fibre – the immediate choice for a horizontal *n*-frame field on *GL*(*M*) is defined by all of the *n*-frames $\mathbf{E}_\mu$ in *H* that are reciprocal to the coframes defined by $\theta^\mu$; one calls these *n* horizontal vector fields the *basic vector fields* of the linear connection we have defined. Since the 1-form $\theta^\mu$ is left-invariant under the action of *GL*(*n*) the basic vector fields are right-invariant. However, since we are not assuming the integrability of the differential system defined by the horizontal sub-bundle we cannot make a statement analogous to (4.1) for horizontal vector fields $\mathbf{E}_\mu$, unless we include a contribution from the vertical $n^2$-frame, which would bring us to the topic of curvature.

First, we need to complete the reciprocal of the $n+n^2$-frame on *GL*(*M*) that is defined by $\{\mathbf{E}_\mu, \mathbf{V}_I\}$. We have already defined the canonical 1-form $\theta^\mu$ to be the reciprocal coframe to the $\mathbf{E}_\mu$. We next introduce the $n^2$-coframe field $\omega^\mu_\nu$ that is reciprocal to the fundamental vector fields $\mathbf{V}_I$, in which we have changed the indexing of the basis vectors for $\mathfrak{gl}(n)$. Hence, the 1-forms $\omega^\mu_\nu$ can also be regarded as a 1-form on *GL*(*M*) that takes its values in $\mathfrak{gl}(n)$ and has the properties that it is $\text{Ad}^{-1}$-equivariant under the action of *GL*(*n*), so for any $h \in GL(n)$:

$$\omega_{\mathbf{e}h} = h^{-1} \omega_\mathbf{e} h. \tag{4.6}$$

(In this expression, we have suppressed the unused indices.)

The coframe $\omega^\mu_\nu$ not only takes each vertical vector space on *GL*(*M*) isomorphically to the vector space $\mathfrak{gl}(n)$, it also has the property that it takes the fundamental vector fields on *GL*(*M*) to their generators in $\mathfrak{gl}(n)$; this follows from the $\text{Ad}^{-1}$-equivariance property.

Moreover, the connection 1-form $\omega_\nu^\mu$ has the property that its annihilating subspaces in $T(GL(M))$ are the horizontal subspaces; i.e.:

$$\omega_\nu^\mu(\mathbf{v}) = 0 \qquad \text{iff} \qquad \mathbf{v} \in H. \tag{4.7}$$

The integrability condition for the exterior differential system in $GL(M)$ defined by $\omega_\nu^\mu = 0$ is then dual to the condition on the Lie brackets of horizontal vector fields, which we shall discuss shortly.

If $\mathbf{e}_\mu: U \to GL(M)$ is a local frame field on $U \subset M$ then the 1-forms $\omega_\nu^\mu$ pull down to a 1-form on $U$ with values in $\mathfrak{gl}(n)$ that we denote by the same symbol. Since the canonical 1-form $\theta^\mu$ also pulls down to a coframe field on $U$ we can express the local 1-form $\omega_\nu^\mu$ as:

$$\omega_\nu^\mu = \Gamma_{\lambda\nu}^\mu \theta^\lambda \tag{4.8}$$

for suitable functions $\Gamma_{\lambda\nu}^\mu$. (Note that this construction makes no sense on $GL(M)$ since the canonical 1-form $\theta^\mu$ and the connection 1-form $\omega_\nu^\mu$ annihilate complementary subspaces of $T(GL(M))$.)

Since $T(GL(M))$ is assumed to be given a direct sum decomposition, any tangent vector $\mathbf{v}$ to $GL(M)$ can be uniquely decomposed into a sum:

$$\mathbf{v} = \mathbf{v}_H + \mathbf{v}_V, \tag{4.9}$$

which defines two canonical projection maps:

$$H: T(GL(M)) \to H, \quad \mathbf{v} \mapsto \mathbf{v}_H, \tag{4.10a}$$
$$V: T(GL(M)) \to V, \quad \mathbf{v} \mapsto \mathbf{v}_V, \tag{4.10b}$$

which can be thought of as either maps between bundles or the maps between corresponding fibres at a chosen frame.

One can easily express the projection maps $H$ and $V$ in terms of the basic vector fields $\mathbf{E}_\mu$, the fundamental vector fields $\mathbf{V}_\nu^\mu$, and their reciprocal coframe fields $\theta^\mu$ and $\omega_\nu^\mu$:

$$H = \theta^\mu \otimes \mathbf{E}_\mu, \qquad V = \omega_\nu^\mu \otimes \mathbf{V}_\mu^\nu. \tag{4.11}$$

An aspect of the map $V$ that will be generalized in a later section is the fact that it can also be regarded as a horizontal 1-form on $GL(M)$ with values in the vector bundle $V$.

We define the *torsion* 2-form of a linear connection on $GL(M)$ – as defined by the horizontal complement of $V$ – to be the 2-form on $GL(M)$ with values in $\mathbb{R}^n$ that is defined by evaluating $d\theta^\mu$ on the horizontal parts of the vector fields:

$$\Theta^\mu(\mathbf{v}, \mathbf{w}) = d\theta^\mu(\mathbf{v}_H, \mathbf{w}_H). \tag{4.12}$$

Similarly, we define the *curvature* 2-form $\Omega^\mu_\nu$ for the connection to be the 2-form on $GL(M)$ with values in $\mathfrak{gl}(n)$ that is defined by evaluating $d\omega^\mu_\nu$ on the horizontal parts:

$$\Omega^\mu_\nu(\mathbf{v},\mathbf{w}) = d\omega^\mu_\nu(\mathbf{v}_H, \mathbf{w}_H). \tag{4.13}$$

A standard result of differential geometry in its post-Cartan formulation (see [**6**, **10**, **11**]) is that these definitions imply the *Cartan structure equations:*

$$d\theta^\mu = -\omega^\mu_\nu \wedge \theta^\nu + \Theta^\mu \tag{4.14a}$$
$$d\omega^\mu_\nu = -\omega^\mu_\kappa \wedge \omega^\kappa_\nu + \Omega^\mu_\nu. \tag{4.14b}$$

When we write them in this form, we immediately see how they relate to the integrability of the exterior differential systems defined by $\theta^\mu = 0$ and $\omega^\mu_\nu = 0$. It helps to know that the 2-forms $\Theta^\mu$ and $\Omega^\mu_\nu$ can be written in terms of the canonical 1-form $\theta^\mu$ as:

$$\Theta^\mu = \tfrac{1}{2} T^\mu_{\kappa\nu}\, \theta^\kappa \wedge \theta^\nu, \qquad \Omega^\mu_\nu = \tfrac{1}{2} R^\mu_{\nu\kappa\lambda}\, \theta^\kappa \wedge \theta^\lambda, \tag{4.15}$$

for suitable functions $T^\mu_{\kappa\nu}, R^\mu_{\nu\kappa\lambda}$ on $GL(M)$. This makes it clear that the integrability of the system $\theta^\mu = 0$, whose integral submanifolds are the fibres of $GL(M)$, follows from the fact that (4.14a) can be locally factored into:

$$d\theta^\mu = -\tfrac{1}{2}(\Gamma^\mu_{[\kappa\nu]} - T^\mu_{\kappa\nu})\,\theta^\kappa \wedge \theta^\nu = -\tfrac{1}{2} c^\mu_{\kappa\nu}\, \theta^\kappa \wedge \theta^\nu, \tag{4.16}$$

in which $\Gamma^\mu_{[\kappa\nu]}$ represents the anti-symmetric part of the array of indices. Hence, Frobenius's theorem gives the integrability. This formula also has the consequence that locally, one has:

$$T^\mu_{\kappa\nu} = \Gamma^\mu_{[\kappa\nu]} + c^\mu_{\kappa\nu}. \tag{4.17}$$

Hence, in a holonomic frame field the torsion vanishes iff the components of the connection 1-form are symmetric in their lower indices.

Now, the situation is quite different for the complementary exterior differential system $\omega^\mu_\nu = 0$ since the two terms in (4.14b) do not have a common factor of $\omega^\mu_\nu$. However, if $\Omega^\mu_\nu = 0$ then the structure equation does give the integrability of this system. Hence, one can regard the curvature of the connection as the obstruction to the integrability of the horizontal sub-bundle. A sufficient, but not necessary condition, for the vanishing of the curvature of a connection is that $M$ be parallelizable and that the connection is the one that makes a chosen global frame field parallel.

The integrability condition for $H$ that is dual to the second structure equation is expressed by the commutation relations of the basic horizontal vector fields:

$$[\mathbf{E}_\mu, \mathbf{E}_\nu] = -\Theta^\lambda(\mathbf{E}_\mu, \mathbf{E}_\nu)\, \mathbf{E}_\lambda - \Omega^\kappa_\lambda(\mathbf{E}_\mu, \mathbf{E}_\nu)\, \mathbf{V}^\lambda_\kappa. \tag{4.18}$$

Clearly, it is the term that involves the curvature 2-form that obstructs the closure of this Lie algebra.

To complete the commutation relations for the adapted frame field $\{\mathbf{E}_\mu, \mathbf{V}_\nu^\mu\}$, we note that since the basic vector fields $\mathbf{E}_\mu$ are right-invariant under the action of $GL(n)$ on frames and the $\mathbf{V}_\nu^\mu$ generate one-parameter subgroups of $GL(n)$, one necessarily has:

$$[\mathbf{E}_\mu, \mathbf{V}_\lambda^\kappa] = -L_{\mathbf{V}_\lambda^\kappa}\mathbf{E}_\mu = 0, \qquad \text{for all } \kappa, \lambda, \mu. \tag{4.19}$$

Since the adapted (global) frame field $\{\mathbf{E}_\mu, \mathbf{V}_\nu^\mu\}$ is clearly semi-holonomic, it is interesting to observe how one might represent it locally in terms of a holonomic frame field that is adapted to the foliation defined by the fibres of $GL(M)$. In order to define such an adapted local holonomic frame field, we start with a local frame field $\mathbf{e}_\mu: U \to GL(M)$, $x \mapsto \mathbf{e}_\mu(x)$. The union of all the fibres of $GL(M)$ over $U$, which we denote by $GL(U)$, has a tangent bundle $T(GL(U))$ that splits into $H(U) \oplus V(U)$ when we have chosen a connection on $GL(M)$.

This tangent bundle can be spanned by the adapted semi-holonomic frame field $\{\mathbf{E}_\mu, \tilde{\mathbf{E}}_I, I = 1,\ldots,n^2\}$, in which we have changed out notation for the fundamental vector fields on the fibres, for brevity. The reciprocal coframe field on $GL(U)$ is then defined by $\{\theta^\mu, \omega^I\}$.

Since the map $\mathbf{e}_\mu: U \to GL(M)$ is an injective immersion, its differential map $D\mathbf{e}_\mu: T(U) \to T(GL(M))$ represents each tangent space $T_x(U)$ as a subspace of $T_{\mathbf{e}(x)}(GL(U))$ that we abbreviate by $\hat{T}_{\mathbf{e}(x)}(U)$, and think of as the "lift" of $T_x(U)$ by $\mathbf{e}_\mu$. We can further extend these subspaces to all of the frames in $GL(U)$ by right translation. Furthermore, one sees that the spaces $\hat{T}_\mathbf{e}(U)$ are all complementary to the vertical subspaces. Hence, we can represent $T(GL(U))$ as the Whitney sum $\hat{T}(U) \oplus V$.

Because the differential is an isomorphism onto, we can also lift tangent vectors to $U$ and, in particular, tangent frames. If we suppose that $(U, x^\mu)$ is a coordinate chart, with $\partial_\mu$ as its natural frame field and $dx^\mu$ as its natural reciprocal coframe field then we can lift to an $n$-frame field on $GL(U)$ that we denote by $\hat{\partial}_\mu$. Similarly, we can pull $dx^\mu$ up an $n$-coframe field that we denote by $\widehat{dx}^\mu$ by using the projection $GL(U) \to U$. We complete the frame field $\hat{\partial}_\mu$ with the same fundamental frame field $\tilde{\mathbf{E}}_I$ on $V(U)$ that we used before and complete the coframe field $\widehat{dx}^\mu$ with the coframe field $\Gamma^I$, which will differ from the $\omega^I$.

The transformation from the adapted holonomic frame field $\{\hat{\partial}_\mu, \tilde{\mathbf{E}}_I\}$ to the adapted semi-holonomic frame field $\{\mathbf{E}_\mu, \tilde{\mathbf{E}}_I\}$ can then take the form:

$$\mathbf{E}_\mu = \hat{\partial}_\mu - N_\mu^I \tilde{\mathbf{E}}_I, \qquad \tilde{\mathbf{E}}_I = \tilde{\mathbf{E}}_I. \tag{4.20}$$

We can identify the components of the matrix $N_\mu^I$ by writing the reciprocal transformation of the coframe fields:

$$\theta^\mu = \widehat{dx}^\mu, \qquad \Gamma^I = \omega^I + N_\mu^I \widehat{dx}^\mu, \tag{4.21}$$

which makes:

$$N_\mu^I = \Gamma^I(\hat{\partial}_\mu) = \Gamma_\mu^I. \tag{4.22}$$

Hence, with a suitable re-indexing of the components $\Gamma_\mu^I$ to give them the form $\Gamma_{\mu\nu}^\kappa$, we see that the matrix $N_\mu^I$ that converts the holonomic frame to the semi-holonomic one is defined by the usual components of the connection 1-form $\omega$ relative to the holonomic frame. This is because one finds that in the decomposition of a lifted vector field $\hat{\mathbf{v}} = v^\mu \hat{\partial}_\mu \in T_{\mathbf{e}}(GL(U))$ into $\hat{\mathbf{v}}_H + \hat{\mathbf{v}}_V$, the vertical part of $\mathbf{v}$ becomes:

$$\hat{\mathbf{v}}_V = \omega^I(\hat{\mathbf{v}})\tilde{\mathbf{E}}_I = v^\mu (\Gamma^I(\hat{\partial}_\nu) \widehat{dx}^\nu (\hat{\partial}_\mu))\tilde{\mathbf{E}}_I = v^\mu \Gamma_\mu^I \tilde{\mathbf{E}}_I, \tag{4.23}$$

which makes the horizontal part equal to:

$$\hat{\mathbf{v}}_H = \hat{\mathbf{v}} - \hat{\mathbf{v}}_V = v^\mu \hat{\partial}_\mu - v^\mu \Gamma_\mu^I \tilde{\mathbf{E}}_I. \tag{4.24}$$

The choice of a horizontal sub-bundle of $T(GL(M))$ allows one to define a special class of curves in $GL(M)$, namely *horizontal curves*; i.e., ones whose tangent vectors are horizontal. If $\rho: (-\varepsilon, +\varepsilon) \to GL(M)$, $\tau \mapsto \rho(\tau)$ is a smooth curve and $\dot{\rho}(\tau)$ is its velocity vector field then this condition can be expressed by the equation:

$$\omega_\nu^\mu(\dot{\rho}(\tau)) = 0 \qquad \text{for all } \tau \in (-\varepsilon, +\varepsilon), \tag{4.25}$$

since the $\omega_\nu^\mu(\dot{\rho}(\tau))$ represent the components of the vertical part of $\dot{\rho}(\tau)$ relative to the fundamental frame field on $V$.

Since a curve in $GL(M)$ will project onto a – possibly degenerate – curve in $M$ one can think of a curve in $GL(M)$ as a frame field along a curve in $M$. Relative to the splitting of $T(GL(M))$ into $H \oplus V$, the motion of a frame along a curve will then consist of two parts: a horizontal motion, which takes it from one point of $M$ to the next, and a vertical motion, which transforms a frame at any point of the curve in $M$ to another frame at the same point. Hence, if there is no vertical motion, one can think of the successive frames along the curve as being parallel along that curve. A horizontal curve in $GL(M)$ then represents the *parallel displacement* of a frame along a curve in $M$.

By standard arguments (cf., e.g., Kobayashi and Nomizu [**10**]), we find that a parallel frame field $\mathbf{e}_\mu(\tau)$ along a curve $\gamma(\tau)$ is a solution to the following system of linear, time-varying, ordinary differential equations:

$$\frac{d\mathbf{e}_\mu}{d\tau} = \omega_\mu^\nu(\mathbf{v})\mathbf{e}_\nu. \qquad (4.26)$$

As the solutions to this system will generally exist only for some finite proper time interval around $\tau = 0$, one sees the root of the fact that generally the parallel translation of frames along a curve is not generally defined for the entire curve, but only a neighborhood of any of its points.

A curve $\gamma(\tau)$ in $GL(M)$ is *vertical* iff its velocity vector field is a vertical vector for every $\tau$. Consequently, such a curve must lie in a single fibre of $GL(M)$. One can then regard such as curve as describing a one-parameter family of linear transformations of an initial frame at a point of $M$ into other frames at that same point. Such a curve then satisfies the equations:

$$\theta^\mu(\dot\gamma(\tau)) = 0, \qquad \text{for all } \tau. \qquad (4.27)$$

One can relate the concepts of the torsion and curvature of a connection on $GL(M)$ by looking at what happens when one parallel translates an affine frame $\mathbf{e}_\mu$ around an infinitesimal loop at a point $x_0 \in M$. In general, it will have been transformed in the process: The values of the curvature 2-form $\Omega_\nu^\mu$ define an infinitesimal linear transformation of the frame while the components of the torsion 2-form $\Theta^\mu$ describe an infinitesimal translation of its origin in the affine tangent space to $x_0$. It is useful in what follows to think of the translation as an infinitesimal horizontal translation and the linear transformation as an infinitesimal vertical translation. In general, each loop will be associated with a different infinitesimal affine transformation, but ultimately all of the resulting transformations will define a subgroup of the affine group in $n$ dimensions that one calls the *holonomy group* of the connection.

Before we move on to the generalization of this section, we briefly summarize the most relevant points:
  a. A linear connection on $GL(M)$ is a splitting of its tangent bundle into a vertical sub-bundle, which is integrable, and a horizontal complement, which does not have to be.
  b. $GL(M)$ is parallelizable. A global frame field that is adapted to the splitting is defined by the basic horizontal vector fields and the fundamental vertical vector fields. The reciprocal coframe field is defined by the canonical 1-form on $GL(M)$ and the connection 1-form.
  c. The integrability conditions for the two sub-bundles can be expressed in terms of the commutation relations for the respective vector fields or the respective exterior derivatives of the reciprocal 1-forms. In either event, the obstruction to the integrability of the horizontal sub-bundle is the non-vanishing of the curvature 2-form.
  d. The splitting $H \oplus V$ distinguishes two classes of curves in $GL(M)$: horizontal curves and vertical curves.

## 5. Almost-product structures.

When a manifold $M$ is expressible as a product manifold $L \times N$ this implies that there is a Whitney sum decomposition of the tangent bundle as $T(M) = T(L) \oplus T(N)$. Conversely, one can start with such a decomposition $T(M) = H(M) \oplus V(M)$ on $M$ and examine the conditions under which this Whitney sum decomposition, which one calls an *almost-product* structure on $M$, is the result of a product structure on the manifold itself. (For some of the basic terminology and geometry of almost-product structures, see Gray [**12**].)

Since the vector spaces of sections of the two bundles $T(L)$ and $T(N)$ define Lie algebras, it is clearly necessary that the sections of $H(M)$ and $V(M)$ must define sub-algebras of $\mathfrak{X}(M)$. As this is precisely the condition that the sub-bundles $H(M)$ and $V(M)$ be integrable, one sees that one is dealing with a problem in the integrability of differential systems when one wishes to establish whether an almost-product structure is due to a product structure. However, although an integrable almost-product structure defines transversal foliations that locally define coordinate charts the make the manifold look like a product manifold they do not have to piece together into a global product structure. For instance, the leaves of either foliation might not all be diffeomorphic.

It is clear that an almost-product structure induces a corresponding Whitney sum decomposition of $T^*(M) = H^*(M) \oplus V^*(M)$ if one maps $H(M) \oplus V(M)$ to $T(M)$ isomorphically by the identity map and pulls back linear functionals on the tangent spaces by way of the identity map. In more practical terms, if $\alpha \in T^*(M)$ and $\mathbf{v} \in T_x(M)$ then one sets:

$$\alpha(\mathbf{v}) = \alpha(\mathbf{v}_H) + \alpha(\mathbf{v}_V) \equiv \alpha_H(\mathbf{v}) + \alpha_V(\mathbf{v}) = (\alpha_H + \alpha_V)(\mathbf{v}) \,. \tag{5.1}$$

When $M$ is given such an almost product structure, the canonical constructions that one immediately defines are based in the fact that any tangent vector $\mathbf{v} \in T_x(M)$ can be uniquely expressed as $\mathbf{v}_H + \mathbf{v}_V$ with $\mathbf{v}_H \in H_x(M)$ and $\mathbf{v}_V \in V_x(M)$. This defines two canonical projections:

$$H_x: T_x(M) \to H_x(M), \quad \mathbf{v} \mapsto \mathbf{v}_H, \qquad V_x: T_x(M) \to V_x(M), \quad \mathbf{v} \mapsto \mathbf{v}_V, \tag{5.2}$$

If $I_x$ represents the identity map on $T_x(M)$ then one has:

$$I_x = H_x + V_x, \qquad V_x = I_x - H_x, \tag{5.3}$$

so, in effect, it is sufficient to define either of the two maps to define the other.

The maps:

$$I: M \to T^*(M) \otimes T(M), \qquad x \mapsto I_x, \tag{5.4a}$$
$$H: M \to T^*(M) \otimes H(M), \qquad x \mapsto H_x, \tag{5.4b}$$
$$V: M \to T^*(M) \otimes V(M), \qquad x \mapsto V_x, \tag{5.4c}$$

define three fields of endomorphisms of the bundle $T(M)$. The endomorphisms $H$ and $V$ satisfy the axioms for complementary linear projections:

$$H^2 = H, \qquad V^2 = V, \qquad HV = VH = 0, \qquad I = H + V\,. \tag{5.5}$$

We can also define another field of endomorphisms by:

$$F = H - V. \tag{5.6}$$

By the rules (5.5), it is seen to be an idempotent:
$$F^2 = I. \tag{5.7}$$

Hence, its eigenvalues are $\pm 1$; the +1 eigenbundle is $H(M)$ and the $-1$ eigenbundle is $V(M)$.

Conversely, one can define a field $F$ of idempotent endomorphisms of $T(M)$ to be an almost-product structure on $M$ and reconstruct the Whitney sum decomposition $H(M) \oplus V(M)$ from the eigenbundles of $F$. Note that the Hodge $*$ isomorphism on a $2n$-dimensional Riemannian manifold has this effect on bundle of $n$-forms. By contrast, if $*^2 = -I$ then one has not defined an almost-product structure, but an *almost-complex* structure, which is what happens with the middle-dimensional forms on an even-dimensional Lorentzian manifold. In such a case, the eigenbundles are not defined if $M$ is a real manifold since the eigenvalues $\pm i$ are imaginary.

As opposed to a foliation, which distinguishes only one sub-bundle $\mathcal{D}(M)$ of $T(M)$, an almost-product structure distinguishes two complementary sub-bundles. Hence, one can reduce the bundle of linear frame on $M$ to a bundle $GL_{n,m}(M)$ of doubly-adapted frames on $M$, where a doubly-adapted frame $\{\mathbf{e}_i, \mathbf{e}_a\}$ is adapted to both the sub-bundles $H(M)$ and $V(M)$. The structure group for this principal bundle is then $GL(n) \times GL(m)$. The transition matrix from one doubly-adapted frame to another will then be of the form:

$$A = \begin{bmatrix} A^i_j & 0 \\ \hline 0 & A^a_b \end{bmatrix}. \tag{5.8}$$

Since the product of a matrix of the form (5.8) with one of the form (3.7) is of the form (2.2), we can see how the group $GL(m+n; n)$ is the semi-direct product of the group $GL(n) \times GL(m)$ with the group $\mathbb{R}^{mn}$, as represented by the invertible matrices of the form (3.7). Since their quotient is contractible, the groups $GL(m+n; n)$ and $GL(n) \times GL(m)$ have the same homotopy groups.

An important point to be made about this reduction is that, generally, unless both sub-bundles are integrable one cannot always find an atlas of doubly-adapted coordinate systems for $M$ whose coordinate transformations have differential maps with matrices like (5.8). Hence, if we are assuming that $H(M)$ is not necessarily integrable the reduction is not complete, and we must still deal with coordinate transformations whose differentials have matrices that look like (2.2).

One can express the operators $I$, $H$, $V$, and $F$ locally in terms of local frame fields quite simply, as long as they are adapted to the splitting. Suppose $\{\mathbf{e}_i, i = 1, \ldots, m, \mathbf{e}_a, a = m+1, \ldots, m+n\}$ is adapted to the almost-product structure on $U \subset M$, and that $\{\theta^i, \theta^a\}$ is its reciprocal frame field, so one has both $\theta^i(\mathbf{e}_j) = \delta^i_j$ and $\theta^a(\mathbf{e}_b) = \delta^a_b$, with all of the other possibilities vanishing. Locally, one then has:

$$I = \theta^I \otimes \mathbf{e}_I, \tag{5.9a}$$
$$H = \theta^i \otimes \mathbf{e}_i, \tag{5.9b}$$
$$V = \theta^a \otimes \mathbf{e}_a, \tag{5.9c}$$
$$F = \theta^i \otimes \mathbf{e}_i - \theta^a \otimes \mathbf{e}_a. \tag{5.9d}$$

When the vector bundle $T(M)$ is given an almost-product structure, as above, this will imply corresponding decompositions of the various tensor bundles over $M$. As particular cases, we mention the decomposition of the bundle $T^{(2,\,0)}(M) = T(M) \otimes T(M)$ of second-rank contravariant tensors:

$$T^{(0,\,2)} = (H \otimes H) \oplus (H \otimes V) \oplus (V \otimes H) \oplus (V \otimes V), \tag{5.10}$$

in which we have abbreviated the superfluous notations, its sub-bundle $S^{(2,0)}(M) = T(M) \odot T(M)$ of symmetric second rank contravariant tensors:

$$S^{(0,\,2)} = (H \odot H) \oplus (H \odot V) \oplus (V \odot V), \tag{5.11}$$

and the sub-bundle $\Lambda_2(M) = T(M) \wedge T(M)$ of anti-symmetric bivector fields:

$$\Lambda_2 = (H \wedge H) \oplus (H \wedge V) \oplus (V \wedge V). \tag{5.12}$$

Following Cattaneo [1], we introduce the generic notation $P_{HH}$, $P_{HV}$, etc., for projections onto the respective factors of these bundles and indicate the terms in the decomposition of a second-rank tensor by subscripts. For instance, a general second-rank tensor field $T$ on $M$ decomposes into a unique sum of terms:

$$T = T_{HH} + T_{HV} + T_{VH} + T_{VV}, \tag{5.13}$$

in which $T_{HH} = P_{HH}(T)$, $T_{HV} = P_{HV}(T)$, etc.

Although the Lie bracket of vector fields defines an anti-symmetric bilinear map from the $C^\infty(M)$-module $\mathfrak{X}(M) \times \mathfrak{X}(M)$ into $\mathfrak{X}(M)$, one does not, however, necessarily have a decomposition of the Lie algebra $\mathfrak{X}(M)$ into $\mathfrak{X}_{HH}(M) \oplus \mathfrak{X}_{HV}(M) \oplus \mathfrak{X}_{VV}(M)$, as in (5.12), since the Lie brackets do not have to close with the respective sub-modules. Once again, this is really a matter of the integrability of the sub-bundles, which is what we shall discuss next. One can also regard this as being an essential difference between dealing with individual tangent vectors and dealing with tangent vector *fields*, which include more local information at each point in the form of their successive derivatives – or *jets*.

In order to account for Cattaneo's definitions of the longitudinal and transverse derivatives relative to an almost-product structure, one need only consider a differentiable map $f: M \to N$ when $M$ is given an almost product structure $T(M) = H(M) \oplus V(M)$. Since the differential of $f$ is a map $Df: T(M) \to T(N)$ that is linear on the fibres, if we first apply the projection $H$ and compose it with $Df$ to obtain $Df \cdot H$ then we have what he was calling the *transverse derivative* of $f$. Similarly, the composition $Df \cdot V$ gives what was calling the *longitudinal derivative* of $f$. For example, in the 1+3 splitting of the spacetime tangent bundle that we shall discuss later the transverse derivatives are directional derivatives in spatial directions and the longitudinal derivatives are essentially the proper time derivatives.

**6. Nonlinear connections.**
The concept of a nonlinear connection on an $m+n$-dimensional manifold $M$ is a generalization of the aforementioned way of defining a connection on a fibre bundle by way of a splitting of its tangent bundle. One must sacrifice the action of a group and the existence of a fibration of $M$ over some lower-dimensional manifold.

A *nonlinear connection* [**5**, **13**] on $M$ is defined by a Whitney sum splitting $H(M) \oplus V(M)$ of $T(M)$ – i.e., an almost-product structure on $M$ – in which the rank $n$ "vertical" sub-bundle $V(M)$ is required to be integrable, but not necessarily the rank $m$ "horizontal" sub-bundle $H(M)$; we shall think of such an almost-product structure as being *semi-holonomic*.

By singling out one of the two bundles this way, we are suggesting that, in a sense, the integrable bundle is more fundamental, and we shall adopt the terminology of calling $V(M)$ the structural sub-bundle while calling $H(M)$ the transversal sub-bundle. One can then think of $H(M)$ as essentially one representation of the normal bundle to $V(M)$. This way of looking at things makes it more straightforward for us to apply the methodology of the transversal geometry of foliations to the immediate subject (see, [**14-16**]).

In effect, since we are requiring the structural sub-bundle to be integrable, we are really generalizing from a fibration to a foliation, since $M$ will be foliated by the $n$-dimensional integral submanifolds of $V(M)$. However, we have abandoned the local triviality that one expects of a fibration. The existence of a group action has been weakened to the action of the groups Diff($M$; $L$) of diffeomorphisms of $M$ that take any leaf of the foliation to itself, so the leaves of this foliation are not required to be orbits of some more elementary group that acts on $M$ globally.

*6.1. Connections as fields of endomorphisms.*
Recall that the effect of the canonical 1-form $\theta^\mu$ on $GL(M)$ is to map horizontal subspaces in $T(GL(M))$ isomorphically to $\mathbb{R}^n$ and vertical subspaces to 0, while a connection 1-form $\omega^\mu_\nu$ maps vertical subspaces isomorphically to the Lie algebra $\mathfrak{gl}(n)$ and horizontal subspaces to 0. One sees that for an almost-product structure the projection $H$ plays a role that is analogous to $\theta^\mu$, mapping transversal subspaces to vector spaces that are isomorphic to $\mathbb{R}^m$ and structural subspaces to 0, and the projection $V$ plays a role that is analogous to $\omega^\mu_\nu$, except that no Lie algebra has been defined on the vector spaces $V_x$, so the model vector space is simply $\mathbb{R}^n$.

We have, however, defined a Lie algebra on $\mathfrak{X}(V)$ by the assumption of the integrability of the sub-bundle $V(M)$, and we then refer to its elements as *structural vector fields*. Hence, we could generalize $V$ from a field of endomorphisms of $T(M)$ to a 1-form on $T(M)$ with values in the sheaf of Lie algebras whose elements are germs of local structural vector fields. However, we shall not pursue this option, at the moment. It is customary in the literature to refer to *H* and *V* as "vector-valued 1-forms," although it would be more precise to say "1-forms with values in vector bundles." We must point out that defining the field of endomorphisms *V* is essentially a generalization of defining an *Ehresmann connection* [3] on a fibre bundle that has been given no group action to the case of a manifold that also has been given no fibration.

If one does not wish to resort to local expressions that involve a choice of local frame field, one can associate each of the fields *H* and *V*, which one regards as 1-forms on *M* with values in the vector bundles *H(M)* and *V(M)*, with 2-forms that take their values in these respective vector bundles by means of the *Fröhlicher-Nijenhuis bracket* [17]:

$$S_V = -\tfrac{1}{2} [\![V,V]\!], \tag{6.6}$$

in which:

$$[\![V,V]\!](\mathbf{v},\mathbf{w}) = 2([\mathbf{v},\mathbf{w}] + [\mathbf{v}_V,\mathbf{w}_V] - V[\mathbf{v},\mathbf{w}_V] - V[\mathbf{v}_V,\mathbf{w}]); \tag{6.7}$$

The brackets on the right-hand side of (6.7) are simply the Lie brackets of the relevant vector fields. This equation can be put into the following form:

$$[\![V,V]\!](\mathbf{v},\mathbf{w}) = 2(H[\mathbf{v}_V,\mathbf{w}_V] + V[\mathbf{v}_H,\mathbf{w}_H]), \tag{6.8}$$

which is more useful for our purposes.

The bilinearity of the bracket, along with the definitions of *H* and *F*, make it easy to obtain the fact that:

$$[\![V,V]\!] = -[\![H,V]\!] = [\![H,H]\!] = \tfrac{1}{4}[\![F,F]\!]. \tag{6.9}$$

Hence, it is unnecessary for us to discuss any of the other brackets that the splitting *H*⊕*V* defines aside from $[\![V,V]\!]$.

Furthermore, we point out that this bracket vanishes when it is applied to vector fields of mixed types:

$$[\![V,V]\!](\mathbf{v}_V,\mathbf{w}_H) = [\![V,V]\!](\mathbf{v}_H,\mathbf{w}_V) = 0, \tag{6.10}$$

which follows immediately from (6.8).

We give the separate terms of (6.8) their own notations:

$$\Omega(\mathbf{v},\mathbf{w}) = -V[\mathbf{v}_H,\mathbf{w}_H], \qquad \tilde{\Omega}(\mathbf{v},\mathbf{w}) = -H[\mathbf{v}_V,\mathbf{w}_V]. \tag{6.11}$$

and, following Kolar, et al. [18], we call them the *curvature* and *co-curvature 2-forms* of the connection 1-form *V*, respectively.

We can then write (6.8) in the form:

$$\Omega + \tilde{\Omega} = -\tfrac{1}{2}[\![V,V]\!]. \tag{6.12}$$

When this sort of bracket was applied in the context of an almost-complex structure, the resulting tensor field was referred to as the *Fröhlicher-Nijenhuis torsion*. It non-vanishing was the obstruction to the integrability of the almost-complex structure. However, since it is curvature that obstructs the integrability of the horizontal sub-bundle of a fibre bundle that has been given a connection, it might also be correct to call it a curvature.

The integrability condition for $V(M)$ takes the form:

$$\tilde{\Omega} = 0 . \tag{6.13}$$

Since we have required that our vertical sub-bundle be integrable, this is tautological.

The integrability condition for $H(M)$ takes the form:

$$\Omega = 0. \tag{6.14}$$

This condition is not necessarily satisfied in the general case, but for a semi-holonomic almost-product structure, one does have:

$$\Omega = -\tfrac{1}{2}[\![V,V]\!]. \tag{6.15}$$

We define the *torsion 2-form* of $H$ and *co-torsion 2-form* of $V$ to be:

$$\Theta(\mathbf{v}, \mathbf{w}) = -H[\mathbf{v}_H, \mathbf{w}_H] , \qquad \tilde{\Theta}(\mathbf{v}, \mathbf{w}) = -V[\mathbf{v}_V, \mathbf{w}_V], \tag{6.16}$$

and obtain, as a consequence:

$$[\mathbf{v}_H, \mathbf{w}_H] = -\Theta(\mathbf{v}_H, \mathbf{w}_H) - \Omega(\mathbf{v}_H, \mathbf{w}_H), \tag{6.17a}$$
$$[\mathbf{v}_V, \mathbf{w}_V] = -\tilde{\Theta}(\mathbf{v}_V, \mathbf{w}_V) , \tag{6.17b}$$

We shall discuss the nature of the bracket $[\mathbf{v}_H, \mathbf{v}_V]$ later.

Equation (6.17a) is formally consistent with the corresponding expression (4.18) for linear connections, and we have incorporated the vanishing of the co-curvature into (6.17b). Recall that the torsion of a linear connection is not an obstruction to the integrability of anything, but only to the commutativity of the horizontal vector fields.

Since (6.15) suggests that the effect of the operator $-\tfrac{1}{2}[\![V,\bullet]\!]$ is analogous to that of a covariant exterior derivative on $k$-forms with values in the horizontal vector fields, we obtain the Bianchi identities for torsion and curvature, in a sense:

$$0 = [\![V,\Theta]\!] = [\![H,\Omega]\!] = -[\![V,\Omega]\!], \tag{6.18a}$$
$$0 = [\![V,\Omega]\!] = -\tfrac{1}{2}[\![V,[\![V,V]\!]]\!] . \tag{6.18b}$$

These follow from the graded Jacobi identity for the Fröhlicher-Nijenhuis bracket (see [**17**, **18**]). Note that (6.18a) differs in form from the usual Bianchi identity.

*6.2. Local expressions in adapted frame fields.*

First, we define an adapted frame field $\{\mathbf{e}_i, \mathbf{e}_a\}$ on an open subset $U \subset M$. Hence, the $m$-frame $\mathbf{e}_i$ spans $H(U)$ and the $n$ frame $\mathbf{e}_a$ spans $V(U)$. If $\{\theta^i, \theta^a\}$ is the reciprocal coframe field to $\{\mathbf{e}_i, \mathbf{e}_a\}$ then one also has that the $\theta^i$ span $H^*(U)$ and the $\theta^a$ span the $V^*(U)$; similarly, the $\theta^i$ annihilate $V$ and the $\theta^a$ annihilate $H$.

We can then express the various fields of endomorphisms that were defined in (5.4a, b, c), as well as the field $F$, in terms of the adapted frame field as in (5.9a-d). One notes that as long as there is a unique relationship between the frame field $\{\mathbf{e}_i, \mathbf{e}_a\}$ and the coframe field $\{\theta^i, \theta^a\}$, it is no loss of generality to deal with $\{\theta^i, \theta^a\}$ as the representation of the field of endomorphisms $I$, $\theta^i$ as the representation of $H$, and $\theta^a$ as the representation of $V$.

So far, these constructions seem to have only an algebraic character that suggests a relationship to the manifold $M$ only by way of its dimension. The point at which they take on a more geometrical significance is when one differentiates them and decomposes the resulting derivatives into structural and transversal parts.

The key issue with the Fröhlicher-Nijenhuis bracket is the way that the Lie algebra of vector fields on $M$ behaves with respect to the almost-product structure. We have required that the vertical vector fields form a Lie sub-algebra, so there must be structure functions $c_{ab}^d$ on $U$ such that:

$$[\mathbf{e}_a, \mathbf{e}_b] = c_{ab}^d \mathbf{e}_d . \tag{6.19}$$

As a consequence, we are assuming that:

$$c_{ab}^i = 0 . \tag{6.20}$$

However, since we have not required integrability of the horizontal sub-bundle the basic horizontal vector fields must satisfy, more generally:

$$[\mathbf{e}_i, \mathbf{e}_j] = c_{ij}^k \mathbf{e}_k + c_{ij}^a \mathbf{e}_a , \tag{6.21}$$

for appropriate structure functions $c_{ij}^k, c_{ij}^a$.

The remaining structure functions can be obtained from:

$$[\mathbf{e}_i, \mathbf{e}_a] = c_{ia}^k \mathbf{e}_k + c_{ia}^b \mathbf{e}_b . \tag{6.22}$$

If the frame $\{\mathbf{e}_i, \mathbf{e}_a\}$ is obtained from a holonomic frame by the action of $GL(m) \times GL(n)$ then these latter structure functions must vanish. However, holonomic adapted frame fields are essentially ones that can be integrated into adapted coordinate frame fields and we are not assuming this sort of integrability. If $\{\mathbf{e}_i, \mathbf{e}_a\}$ is semi-holonomic then, from (3.2), the $c_{ia}^k$ will vanish, but not necessarily the $c_{ia}^b$, in general.

We obtain the structure equations if we start with the Ansätze:

$$\Theta^i(\mathbf{v}, \mathbf{w}) = d\theta^i(\mathbf{v}_H, \mathbf{w}_H), \qquad \Omega^a(\mathbf{v}, \mathbf{w}) = d\theta^a(\mathbf{v}_H, \mathbf{w}_H), \tag{6.23a}$$

$$\tilde{\Theta}^a(\mathbf{v}, \mathbf{w}) = d\theta^a(\mathbf{v}_V, \mathbf{w}_V), \qquad \tilde{\Omega}^i(\mathbf{v}, \mathbf{w}) = d\theta^i(\mathbf{v}_V, \mathbf{w}_V) = 0, \qquad (6.23b)$$

which correspond to (6.11) and (6.16), and expand:

$$d\theta^i = -\tfrac{1}{2} c^i_{jk} \theta^j \wedge \theta^k - \tfrac{1}{2} c^i_{ja} \theta^j \wedge \theta^a - \tfrac{1}{2} c^i_{ab} \theta^a \wedge \theta^b, \qquad (6.24a)$$

$$d\theta^a = -\tfrac{1}{2} c^a_{jk} \theta^j \wedge \theta^k - \tfrac{1}{2} c^a_{jb} \theta^j \wedge \theta^b - \tfrac{1}{2} c^a_{bc} \theta^b \wedge \theta^c. \qquad (6.24b)$$

If we incorporate the information about the structure functions that follows from the integrability assumptions then this gives:

$$\Theta^i = -\tfrac{1}{2} c^i_{jk} \theta^j \wedge \theta^k, \quad \Omega^a = -\tfrac{1}{2} c^a_{jk} \theta^j \wedge \theta^k, \qquad (6.25a)$$

$$\tilde{\Theta}^a = -\tfrac{1}{2} c^a_{bc} \theta^b \wedge \theta^c, \quad \tilde{\Omega}^i = -\tfrac{1}{2} c^i_{ab} \theta^a \wedge \theta^b = 0. \qquad (6.25b)$$

Hence, if $\{\mathbf{e}_i, \mathbf{e}_a\}$ is semi-holonomic then of these only $\Omega^a$ is non-vanishing.

One should be careful when computing with these expressions, since although, for instance, it *is* true that:

$$\Theta^i(\mathbf{e}_j, \mathbf{e}_k) = -\theta^i[\mathbf{e}_j, \mathbf{e}_k] = c^i_{jk}, \qquad (6.26)$$

it is *not* true that:

$$\Theta^i(\mathbf{X}, \mathbf{Y}) = -\theta^i[\mathbf{X}, \mathbf{Y}], \qquad (6.27)$$

more generally.

This is because although both sides of the equations involve expressions that bilinear over the real *vector space* of vector fields on $U$, only the 2-forms $\Theta^i$ are bilinear over the $C^\infty(U)$-*module* of vector fields on $U$. By direct calculation, one sees that, in fact:

$$\Theta^i(X^j \mathbf{e}_j, Y^k \mathbf{e}_k) = X^j Y^k \, \Theta^i(\mathbf{e}_j, \mathbf{e}_k) = c^i_{jk} X^j Y^k, \qquad (6.28)$$

but:

$$-\theta^i[X^j \mathbf{e}_j, Y^k \mathbf{e}_k] = -(\mathbf{X} Y^j - \mathbf{Y} X^j) + c^i_{jk} X^j Y^k. \qquad (6.29)$$

Hence, the structure functions for a given local frame field do not contain all of the information that it takes to completely determine the Lie algebra of vector fields on their domain of definition, unless one remembers to add in the extra term that appears in (6.29), but not (6.28).

*6.3. The Bott connection.*

One must note that we have not accounted for the coefficients of the mixed exterior products namely, $c^i_{ja}$ and $c^a_{jb}$, in (6.22), which can also be regarded as:

$$[\mathbf{e}_i, \mathbf{e}_a] = H[\mathbf{e}_i, \mathbf{e}_a] + V[\mathbf{e}_i, \mathbf{e}_a]. \qquad (6.30)$$

The first term in (6.30) is a generalization of the *Bott connection* for a foliation (see, Kamber and Tondeur [**14**]). In its simplest form, it allows one to define the covariant derivative of a section of the normal bundle, which, in the present case, will be a transversal vector field **w**, in the direction of a vector field tangent to the foliation, which we represent by a structural vector field **v**, by way of:

$$\overset{o}{\nabla}_{\mathbf{v}} \mathbf{w} = H[\mathbf{v}, \mathbf{w}] = H(L_{\mathbf{v}}\mathbf{w}). \tag{6.31}$$

Hence, one can think of it as the transversal part of the Lie derivative of **w** along the flow of **v**.

When this applied to the members of an adapted local frame field one gets:

$$\overset{o}{\nabla}_{\mathbf{e}_a} \mathbf{e}_i = \theta^j([\mathbf{e}_a, \mathbf{e}_i])\mathbf{e}_j = c_{ai}^j \mathbf{e}_j. \tag{6.32}$$

This will vanish if the frame field is semi-holonomic. Although the mixed structure functions $c_{aj}^i$ will vanish for a semi-holonomic local frame field, the $c_{jb}^a$ will not, in general.

The fact that the Bott connection is flat follows from the definition of curvature in the form:

$$\overset{o}{\Omega}(\mathbf{u}, \mathbf{v})\mathbf{w} = \overset{o}{\nabla}_{\mathbf{u}} \overset{o}{\nabla}_{\mathbf{v}} \mathbf{w} - \overset{o}{\nabla}_{\mathbf{v}} \overset{o}{\nabla}_{\mathbf{u}} \mathbf{w} - \overset{o}{\nabla}_{[\mathbf{u},\mathbf{v}]} \mathbf{w}, \tag{6.33}$$

by way of (6.31), the Jacobi identity, and the assumption that **u**, **v** are structural vector fields, while **w** is a transversal vector field.

*6.4. Horizontal curves.*

We can define horizontal curves on a manifold that is given an almost-product structure $H(M) \oplus V(M)$ in a manner that is analogous to the way that that they were defined for *GL*(*M*). However, a significant difference arises in the fact that since we are not assuming that *M* is fibred over anything, it is not necessary to think in terms of lifts.

Suppose $\gamma: (-\varepsilon, +\varepsilon) \to M$, $\tau \mapsto \gamma(\tau)$ is a curve and $\dot{\gamma}(\tau)$ is its velocity vector field. Then $\gamma$ is horizontal iff $\dot{\gamma}(\tau) \in H_{\gamma(\tau)}(M)$ for all $\tau$, i.e., $V(\dot{\gamma}(\tau)) = 0$ for all $\tau$. If $\{\mathbf{e}_i, \mathbf{e}_a\}$ is an adapted semi-holonomic frame field on $U \subset M$ then this becomes the local condition:

$$\theta^a(\dot{\gamma}(\tau)) = 0 \qquad \text{for all } \tau \in (-\varepsilon, +\varepsilon). \tag{6.34}$$

If $\{U, x^i, x^a\}$ is a local coordinate chart that is adapted to $V(U)$ and $\{\mathbf{e}_i, \mathbf{e}_a\}$ is related to $\{\partial_i, \partial_a\}$ by projection of $\mathbf{e}_i$ onto $H(U)$ then if:

$$\dot{\gamma} = \dot{x}^i \partial_i + \dot{x}^a \partial_a = v^i \mathbf{e}_i \tag{6.35}$$

we can deduce the following system of ordinary differential equations for $\gamma$ from (6.34):

$$\dot{x}^a - V_i^a \dot{x}^i = 0, \qquad V_i^a \equiv \theta^a(\partial_i), \tag{6.36}$$

in which $\{\theta^i, \theta^a\}$ is the reciprocal coframe field to $\{\mathbf{e}_i, \mathbf{e}_a\}$.

*6.5. Orthogonal decompositions.*
Since one often encounters the condition that the two complementary sub-bundles $H(M)$ and $V(M)$ be orthogonal with respect to some metric structure on the $T(M)$, it is interesting to note how the requirement of orthogonality relates to the integrabilility of adapted frame fields. We phrase this as a theorem:

**Theorem:**

Suppose $H(M)$ and $V(M)$ are orthogonal complements in $T(U)$ and $V(M)$ is integrable. If a natural frame field $\{\partial_i, \partial_a\}$ is orthonormal and adapted to $V(U)$ then $H(U)$ must integrable.

**Proof:**
The local frame field $\partial_i$ spans an orthogonal complement to $V_x(U)$ at each $x \in U$. By the uniqueness of orthogonal complements this space must be $H_x(U)$. But that means $H(U)$ can be spanned by a natural frame field, which makes it integrable.

This theorem has some equally interesting restatements that we phrase as a corollary:

**Corollary:**

*a*) Any natural frame field that is adapted to $V(U)$ must be non-orthogonal.

*b*) If a natural frame field is orthonormal and adapted to $V(U)$ then a non-integrable $H(U)$ cannot be the orthogonal complement to $V(U)$.

Statement *a*) affects the way that the metric tensor field can appear in a natural frame field when $H(M)$ is not integrable. If $\{\theta^i, \theta^a\}$ is a doubly-adapted semi-holonomic orthonormal coframe field on $U \subset M$, so [3]:

$$g = g_H + g_V = \eta_{ij}\, \theta^i\, \theta^j + \eta_{ab}\, \theta^a\, \theta^b \tag{6.37}$$

then if the transformation from the chosen coframe field to a natural frame field that is adapted to $V(U)$ takes the form (3.4), the metric will take the form:

$$g = (\eta_{ij} - \eta_{ab} N^a_i N^b_j)\, dx^i\, dx^j - 2\eta_{ab} N^b_i\, dx^i\, dx^a + \eta_{ab}\, dx^a\, dx^a. \tag{6.38}$$

In Vacaru, et al. [**5**], such a metric is called "off-diagonal."

# 7. Physical observers.
In the next section, we shall specialize the foregoing formalism to the case in which $(M, g)$ is a four-dimensional Lorentzian manifold whose pseudo-metric has signature type $(+1, -1, -1, -1)$ and the almost-product structure takes the form of a 1+ 3 splitting $L(M) \oplus \Sigma(M)$ of $T(M)$ into a timelike line field (i.e., a real line bundle) $L(M)$ and a spacelike rank-three sub-bundle $\Sigma(M)$. Most commonly in relativity theory, the spatial sub-bundle is assumed to be the orthogonal complement of $L(M)$ with respect to $g$.

---
[3] We have indicated the symmetrized tensor product in the manner that is customary for metrics.

In order to motivate the mathematical choices, it is useful to first briefly discuss the physical origin of such a splitting. Generally, the timelike line field is tangent to a congruence of timelike curves that is defined by the motion of some distribution of extended matter, such as the streamlines of a fluid motion; i.e., we start with a one-dimensional foliation of $M$. In such a case, one would expect that the splitting was defined only over the support of the mass density distribution, in which case, the manifold $M$ would not represent all of spacetime, but only a world-tube in spacetime. However, it is conceivable that one can adapt the methodology to the case of cosmological models and the time evolution of spatial geometry due to the expansion of the universe – otherwise known as geometrodynamics, – in which case the manifold $M$ might represent all of spacetime. However, in such a case one might have to regard the presence of singularities as obstructions to the complete integrability of the spatial sub-bundle. Of course, most cosmological models are based on the assumption that spacetime is not filled a large number of spatially distant masses, but a continuous distribution of "cosmic dust," which brings one back to a hydrodynamical approximation for geometrodynamics. However, one must eventually introduce singularities in order to account for the presence of black holes and the Big Bang.

One thinks of the motion that defines the congruence as an *observer* [**3**, **4**, **19**] and the complementary spatial sub-bundle as the *rest space* for that motion, relative to the Lorentzian structure. There is some merit in generalizing the splitting to a manifold on which no Lorentzian structure has been given, in which case the choice of complementary rest spaces takes on a more projective geometric character (see, Delphenich [20]); however, we shall not pursue that option here.

*7.1. Orientability and time orientability.*
Clearly, it is mostly the character of the motion that defines $L(M)$ that dictates the geometrical and topological nature of the splitting. However, the topology of $M$ can influence it, as well.

In particular, the question of whether $L(M)$ is trivial or not amounts to the question of whether it is orientable or not. Since the structure group of $L(M)$ is $(\mathbb{R}^*, \times)$, which is homotopically equivalent to $\mathbb{Z}_2$, the obstruction to its orientability is a $\mathbb{Z}_2$-cocyle $w_1(L)$ in dimension one, which is not to be confused with the first Stiefel-Whitney class $w_1$ of $M$, which relates to the orientability of $T(M)$. As a consequence, if $M$ is simply connected, $L(M)$ must be trivial. When $L(M)$ is orientable, one calls $M$ *time-orientable*. It is not hard to see that if $T(M)$ and $L(M)$ are both orientable then so is $\Sigma(M)$.

One can express this fact in terms of unit-volume elements on the various bundles. Suppose $T(M)$ is orientable and oriented, with a unit-volume element given by a globally non-zero 4-form $\mathcal{V}$ whose local representation in terms of a local coframe field $\{\theta^\mu, \mu = 0, \ldots, 3\}$:

$$\mathcal{V} = \theta^0 \wedge \theta^1 \wedge \theta^2 \wedge \theta^3 = \frac{1}{4!} \varepsilon_{\kappa\lambda\mu\nu} \theta^\kappa \wedge \theta^\lambda \wedge \theta^\mu \wedge \theta^\nu, \quad (7.1)$$

and that *L*(*M*) is orientable, with a generating non-zero temporal vector field **t** such that $\theta^\mu(\mathbf{t}) = \delta^\mu_0$. The way that one defines the unit-volume element on Σ(*M*) is by interior product:

$$\mathcal{V}_s = i_\mathbf{t}\mathcal{V} = \theta^1 \wedge \theta^2 \wedge \theta^3 = \frac{1}{3!}\varepsilon_{ijk}\,\theta^i \wedge \theta^j \wedge \theta^k. \tag{7.2}$$

Although we are trying to put off the introduction of a Lorentzian metric on *T*(*M*) until it becomes unavoidable, it is important to point out that – up to homotopy – the very existence of a global line field on a manifold *M* is equivalent to the existence of a Lorentzian structure (see, Markus [**21**], and references in Delphenich [**22**]). If *M* is non-compact then there is no topological obstruction to this, but if *M* is compact then its Euler-Poincaré characteristic must vanish, which is also the necessary and sufficient condition for the existence of a global non-zero vector field. However, there is nothing geometrically canonical about the Lorentzian structure that topology necessitates.

*7.2. Observers as dynamical systems.*
One can gain some intuition for the possible details associated with **t** by regarding it as a dynamical system. Hence, it will either admit a global flow or *M* will admit a covering by local flows. A global flow amounts to an action of the group (ℝ, +), which we shall call the *proper time displacement group* on *M*:

$$\mathbb{R} \times M \to M, \quad (\tau, x) \mapsto x_\tau, \tag{7.3}$$

with **t** the fundamental vector field associated with 0.

There are three basic types of orbits for any action of (ℝ, +) – i.e., three types of integral curves: fixed points, open curves, and closed curves, or *periodic orbits*. If we have assumed that **t** is non-zero, as time-orientability would imply, then we are assuming that there will be no fixed points. The difference between open and closed curves is the difference between an effective action and an ineffective one. A group action is *effective* iff the isotropy subgroup is the identity at all points, where the *isotropy subgroup* at any point is the subgroup of all elements in the transformation group that fix that point. If the action is not effective then the isotropy subgroup might be isomorphic to ℤ, such as the groups $T\mathbb{Z} = \{nT, n \in \mathbb{Z}\}$, for some non-zero period $T \in \mathbb{R}$. In either event, a global flow constitutes an almost-free group action (i.e., the isotropy subgroups are always discrete), so its orbits constitute a foliation for that reason, as well. Since ℝ possesses other additive subgroups besides 0, ℤ, and ℝ – for instance, the rational numbers ℚ – there are other possible orbit types, but they are more pathological in character, and of less physical interest.

It is customary in physics to rule out closed timelike curves on the grounds of causality, especially if one is assuming that the curves are geodesics. However, one could argue that the period *T* might be of cosmological magnitude ( > $10^{10}$ years), and that there is nothing to say that the periods of neighboring curves of the congruence are equal. If one chooses a set of proper-time simultaneous points along various curves – say, the planets of our Solar system – then, unless they all have the same cosmic period (which is not be confused with their planetary periods), when one of them returns to an arbitrarily chosen spacetime point along its integral loop, the other planets would not be in the same spatial configuration, anyway, so one would not actually observe the repetition as a duplication of the spatial state of the entire matter distribution.

Note that time-orientability does not imply the non-existence of closed timelike curves, even though one could embark on such a curve into the future and return to the starting point from the past. A counterexample would be any product spacetime of the form $\Sigma \times S^1$, in which $S^1$ represents the proper-time manifold. A breakdown of time-orientability is more closely associated with periodic orbits such that the velocity vector starts off future-pointing and comes back past-pointing, or with points at which the neighboring vectors of the field **t** are either all future-pointing or all past-pointing.

Just as a non-orientable manifold possesses an orientable covering manifold, similarly, one can define time-orientable covering manifolds for non-time-orientable manifolds and aperiodic covering manifolds for manifolds on which a flow has closed orbits. However, we shall not use these facts any further in this discussion, so we shall not elaborate upon them.

Note that since the fibres of $L(M)$ are one-dimensional, from Frobenius, it must be completely integrable. Hence, $M$ will be foliated by a global congruence of integral curves. Ultimately, the key issues concerning the character of the foliation defined by $L(M)$ in the present context are whether it is a Lie foliation under the action of $(\mathbb{R}, +)$ and whether it is a fibration whose fibres are the orbits.

In order for the foliation to be a Lie foliation, one must satisfy two conditions: there can be no fixed points, since they represent orbits of lower dimension than one, and one must have a global action of $(\mathbb{R}, +)$; i.e., a global flow. Now, the existence of a one-dimensional foliation is not to be confused with the existence of a global flow, because the difference between an integral curve of the foliation and an orbit of a global flow is a matter of parameterization. In general, one must deal with local flows, but if $M$ is compact, the flow will be global, which follows from the fact that one can reduce any open covering by local flows to a finite sub-covering.

In order for the foliation to be a fibration, one must satisfy two conditions: the fibres – i.e., the orbits – must all be diffeomorphic, and there must be a global slice to the foliation − viz., a three-dimensional spatial submanifold that intersects the entire congruence transversally. However, even the existence of a global flow does not imply the existence of a global slice to the flow. Physically, this would amount to a global definition of proper-time simultaneity at some instant – say $\tau = 0$, − which implies that it could be valid only for the observer in question.

*7.3. Types of flows.*
If one assumes that the curves of the congruence represent natural curves then one will generally assume that they are geodesics of some Lorentzian metric $g$, as well. In such a case, one assumes that **t** must satisfy the constraint:

$$0 = \nabla_\mathbf{t} \mathbf{t} = \left(\frac{dt^\mu}{d\tau} + \Gamma^\mu_{\kappa\lambda} t^\kappa t^\lambda\right) \mathbf{e}_\mu, \tag{7.4}$$

relative to the covariant derivative defined by the Levi-Civita connection. The right-hand side of the equation is variously interpreted as the *covariant acceleration* vector field of the congruence or its *geodesic curvature*.

One can also classify the types of flows that $\mathbf{t}$ might define in terms of continuum-mechanical notions. At the root of this classification, one has the fact that an element $a^\mu_\nu$ of the Lie algebra $\mathfrak{gl}(4; \mathbb{R})$, which we regard as a 4×4 real matrix, can be Lorentz polarized into a unique sum:

$$a^\mu_\nu = e^\mu_\nu + \omega^\mu_\nu + \frac{1}{4}\mathrm{Tr}(a)\delta^\mu_\nu, \tag{7.5}$$

in which:

$$e^\mu_\nu = \tfrac{1}{2}(a^\mu_\nu + \tilde{a}^\mu_\nu) \tag{7.6}$$

is the infinitesimal generator of a *Lorentz strain*, which is also referred to as an *infinitesimal shear*, the term:

$$\omega^\mu_\nu = \tfrac{1}{2}(a^\mu_\nu - \tilde{a}^\mu_\nu) \tag{7.7}$$

is the infinitesimal generator of a Lorentz transformation, so $\omega^\mu_\nu \in \mathfrak{so}(3, 1)$, and the last term in (7.5) is the infinitesimal generator of a *dilatation*.

In these expressions, we have introduced the notation:

$$\tilde{a}^\mu_\nu = (\eta a^T \eta)^\mu_\nu \tag{7.8}$$

to describe the *Lorentz adjoint* of a matrix.

When one applies this decomposition to the second-rank tensor field defined by the covariant differential of the covector field $\theta = \theta^0 = i_\mathbf{t} g = t_\mu dx^\mu$ that is metric-dual to $\mathbf{t}$:

$$\nabla \theta = t_{\mu;\,\nu} \partial_\mu \otimes dx^\nu \tag{7.9}$$

one arrives at the following covariant components:

$$e_{\mu\nu} = \tfrac{1}{2}(t_{\mu;\nu} + t_{\nu;\mu}) = \tfrac{1}{2}L_\mathbf{t} g \tag{7.10}$$

for the infinitesimal shear:

$$\omega_{\mu\nu} = \tfrac{1}{2}(t_{\mu;\nu} - t_{\nu;\mu}) = \tfrac{1}{2}(d\theta)_{\mu\nu} \tag{7.11}$$

for the infinitesimal rotation, and:

$$\mathrm{Tr}(\nabla\theta) = t^\mu_{;\,\mu} \tag{7.12}$$

for the infinitesimal dilatation.

From Killing's theorem, $e_{\mu\nu}$ vanishes iff **t** is a *Killing vector field*. It is then an infinitesimal generator of a flow of Lorentzian isometries.

One refers to the 2-form $d\theta$ as the *(kinematical) vorticity* of the flow defined by **t**. One must note that it can be defined independently of the choice of any connection. When the vorticity vanishes, the flow is called *irrotational*.

When the dilatation vanishes, the flow is Lorentz volume-preserving. One then calls it *incompressible*.

*7.4. Decomposition of tensors.*

If we specialize the discussion of the decomposition of tensors for a given 1+3 splitting of $T(M)$, we see that the fact that $L(M)$ is of rank one implies certain simplifications. First, we consider the general case of second rank covariant tensor fields:

$$T^{(0,2)} = (L \oplus \Sigma) \otimes (L \oplus \Sigma) = (\lambda\, \theta \otimes \theta) \oplus (L \otimes \Sigma) \oplus (\Sigma \otimes L) \oplus T^{(0,2)}(\Sigma), \qquad (7.13)$$

in which the first sub-bundle is a real line bundle that consists of all scalar function multiples of the tensor field $\theta \otimes \theta$, the middle two bundles are the space-time and time-space tensors, and the last one $T^{(0,2)}(\Sigma)$ represents the bundle of second rank covariant tensors of purely spatial type. If $\{\theta, \theta^i\}$ is an adapted local coframe field then a general second rank covariant tensor field $T$ can be expressed in the form:

$$T = T_{00}\, \theta \otimes \theta + T_{0i}\, \theta \otimes \theta^i + T_{i0}\, \theta^i \otimes \theta + T_{ij}\, \theta^i \otimes \theta^j. \qquad (7.14)$$

By symmetrizing (7.13), one sees that the bundle of symmetric second-rank tensors admits the decomposition:

$$S^2 = (L \oplus \Sigma) \odot (L \oplus \Sigma) = (\lambda\, \theta \otimes \theta) \oplus (L \odot \Sigma) \oplus S^2(\Sigma), \qquad (7.15)$$

in which $S^2(\Sigma)$ denotes the bundle of symmetric second-rank covariant tensors of purely spatial type, and the other terms are self-explanatory. The local form of such a tensor field with respect to an adapted coframe field is then:

$$g = g_{00}\, \theta^2 + 2g_{i0}\, \theta^i\, \theta + g_{ij}\, \theta^i\, \theta^j, \qquad (7.16)$$

in which $g_{ij}$ is symmetric in its indices. If $L$ is orthogonal to $\Sigma$ then the middle term vanishes.

By anti-symmetrizing (7.13), one can obtain the decomposition for the bundle of 2-forms, and we give the decompositions for the other *k*-forms, for the sake of completeness:

$$\Lambda^1 = L^* \oplus \Sigma^* \qquad = [\theta] \oplus \Lambda^1(\Sigma), \qquad (7.17a)$$
$$\Lambda^2 = (L^* \oplus \Sigma^*) \wedge \Lambda^1 = [\theta] \wedge \Lambda^2(\Sigma) + \Lambda^2(\Sigma), \qquad (7.17b)$$
$$\Lambda^3 = (L^* \oplus \Sigma^*) \wedge \Lambda^2 = [\theta] \wedge \Lambda^2(\Sigma) + \lambda\, \mathcal{V}_s, \qquad (7.17c)$$
$$\Lambda^4 = (L^* \oplus \Sigma^*) \wedge \Lambda^3 = [\theta] \wedge \Lambda^3(\Sigma) = \lambda\, \theta \wedge \mathcal{V}_s. \qquad (7.17d)$$

Hence, the local forms of the respective *k*-forms are:

$$\alpha = \alpha_0 \theta + \alpha_i \theta^i, \tag{7.18a}$$

$$F = F_{0i}\, \theta \wedge \theta^i + \tfrac{1}{2} F_{ij}\, \theta^i \wedge \theta^j, \tag{7.18b}$$

$$H = \tfrac{1}{2} H_{0ij}\, \theta \wedge \theta^i \wedge \theta^j + \frac{1}{3!} H_{ijk}\, \theta^i \wedge \theta^j \wedge \theta^k, \tag{7.18c}$$

$$\lambda \mathcal{V} = \lambda\, \theta \wedge \mathcal{V}_s. \tag{7.18d}$$

Hence, all of the spaces of *k*-forms for *k* = 1, 2, 3 admit a decomposition into a temporal space that takes the form of the image of the linear injection $\Lambda^{k-1} \to \Lambda^k$, $\alpha \mapsto \theta \wedge \alpha$ and another spatial subspace. Of course, by Hodge duality $\Lambda^1$ is isomorphic to $\Lambda^3$, an isomorphism that takes $\theta$ to $*\theta = \mathcal{V}_s$ and the subspace $\Lambda^1(\Sigma)$ to $\theta \wedge \Lambda^2(\Sigma)$.

## 8. Convected frame fields.

We can define a particularly useful class of local frame fields when we are given a local flow for a vector field that we shall call *convected frame fields*. Such frame fields have the property that their structure functions contain geometric information about the nature of the flow, which has the effect of defining the transversal geometry of the flow in terms of the flow itself.

Say we have a system of (proper-) time-varying nonlinear ordinary differential equations on $\mathbb{R}^n$:

$$\frac{dx^i}{d\tau} = v^i(\tau, x^j), \qquad i, j = 1, \ldots, n, \tag{8.1}$$

which is defined by the vector field $\mathbf{v} = v^i \partial_i$; here, we are using a general coordinate system.

The vector field $\mathbf{v}$ has a local flow defined on $U \subset \mathbb{R}^n$ in the form $\Phi: \mathbb{R} \times U \to \mathbb{R}^n$, $(\tau, x_0^i) \mapsto \Phi^i(\tau, x_0^i)$. We can regard the data $x_0^i$ as representing the data of an initial-value problem. Hence, for a given $x_0^i$ at $t = 0$, one can obtain the integral curve through that initial point by means of $x^i(\tau) = \Phi^i(\tau, x_0^i)$. By differentiating this with respect to proper time, we obtain:

$$\frac{dx^i}{d\tau} = \Phi^i_{,0} + \Phi^i_{,j} v_0^j = v^i(\tau, x^j(\tau)), \tag{8.2}$$

in which the comma in the subscript refers to partial differentiation by the relevant coordinate.

We extend $\mathbb{R}^n$ to $\mathbb{R}^{n+1}$ on the left by the additional coordinate $x^0 = \tau$, so the flow $\Phi$ extends to the diffeomorphism of $\mathbb{R} \times U$ onto its image in $\mathbb{R}^{n+1}$: $\Phi^\mu(\tau, x_0^i) = (\tau, \Phi^i(\tau, x_0^i))$. Furthermore, we then have $dx^0/d\tau = 1$. We can then re-write (8.2) as a matrix equation:

$$v^\mu(\tau, x_0^i) = \begin{bmatrix} 1 \\ \hline v^i(\tau) \end{bmatrix} = \begin{bmatrix} 1 & | & 0 \\ \hline \Phi^i_{,0} & | & \Phi^i_{,j} \end{bmatrix} \begin{bmatrix} 1 \\ \hline v_0^i \end{bmatrix} = \Phi^\mu_{,\nu}(\tau, x_0^i) v_0^\nu, \tag{8.3}$$

with $v_0^\mu \equiv v^\mu(0, x_0^i)$. (We have suppressed the explicit inclusion of $x_0^i$ elsewhere in the notation for the sake of brevity.) One can think of the $\Phi^i_{,0}$ components as representing the velocity of the coordinate system $(U, x^i)$ with respect to the given vector field $\mathbf{v} = \partial_0 + v^i \partial_i$. Note that the matrix $\Phi^\mu_{,\nu}$, which is necessarily invertible since $\Phi^\mu(\tau, x_0^i)$ is a diffeomorphism for all $\tau$, is an element of $GL(n+1; n)$, the group of all invertible $(n+1) \times (n+1)$ real matrices that take the affine hyperplane $v^0 = 1$ to itself.

We can further simplify the problem by choosing a coordinate system $(U, x^i)$ that is also adapted to $\mathbf{v} = \partial_0$, which then makes $v^0(\tau) = 1$, $v^i(\tau) = 0$, and also $\Phi^i_{,0} = 0$. This amounts to a co-moving coordinate frame field.

The matrix of $D\Phi$ then takes the form:

$$\Phi^\mu_{,\nu} = \begin{bmatrix} 1 & | & 0 \\ \hline 0 & | & \Phi^i_{,j} \end{bmatrix}. \tag{8.4}$$

Next, we define a convected frame field along the flow of $\mathbf{v}$ by choosing an initial frame field on the $\tau = 0$ slice of $U$ – such as $\mathbf{e}_i(0, x^j) = \partial_i$ – and then mapping it to the other points of the image of $\Phi$ by way of $\Phi^\mu_{,\nu}$. Explicitly, this gives:

$$\mathbf{e}_0(\tau) = \partial_0, \qquad \mathbf{e}_i(\tau) = \Phi^j_{,i} \partial_j. \tag{8.5}$$

Relative to such a frame field, the velocity components will always be equal to their initial values $v_0^\mu = \delta_0^\mu$; i.e.:

$$\mathbf{v} = \mathbf{e}_0. \tag{8.6}$$

which essentially transfers the information that was contained in the velocity vector field into the frame field.

The non-zero structure functions of this local frame field are obtained immediately:

$$c_{0i}^j = \Phi^k_{,i,0} \tilde{\Phi}^j_{,k}, \qquad c_{ij}^k = \Phi^l_{,[i} \Phi^m_{,j],l} \tilde{\Phi}^k_{,m}. \tag{8.7}$$

In these expressions, the tilde over a matrix signifies that we are referring to the inverse of the matrix, and the square brackets around indices mean that the indices concerned are anti-symmetrized.

One way of looking at the structure functions of the frame field $\mathbf{e}_\mu$, which we express in the general form:

$$c_{\mu\nu}^\kappa = \Phi^\rho_{,[\mu} \Phi^\sigma_{,\nu],\rho} \tilde{\Phi}^\kappa_{,\sigma}, \tag{8.8}$$

is that they are the components of the torsion 2-form associated with the flat connection $a^\kappa_{\mu\nu} dx^\kappa$ that makes $\mathbf{e}_\mu$ parallel:

$$c^\kappa_{\mu\nu} = a^\kappa_{[\mu\nu]}, \qquad a^\kappa_{\mu\nu} \equiv \Phi^\rho_{,\mu} \Phi^\sigma_{,\nu,\rho} \tilde{\Phi}^\kappa_{,\sigma}. \tag{8.9}$$

Equations (8.8) clearly show that the structure functions of the frame field depend solely upon a certain combination of first and second derivatives of the flow, which also relate to the differential of the vector field $\mathbf{v}$.

In particular, if one differentiates (8.3) then one sees that:

$$v^\mu_{,\nu} = \Phi^\mu_{,\lambda,\nu} v^\lambda_0 = \Phi^\mu_{,\nu,0}. \tag{8.10}$$

However, these are the components of $D\mathbf{v}$ with respect to the $\partial_\mu$ frame field. If one transforms to the $\mathbf{e}_\mu$ frame field then one finds that the components of $D\mathbf{v}$ take the form:

$$\bar{v}^\mu_{,\nu} = a^\mu_{0\nu} = \Phi^\kappa_{,\nu,0} \tilde{\Phi}^\mu_{,\kappa}. \tag{8.11}$$

This has the form of the components of a generalized "angular" velocity of the $\mathbf{e}_\mu$ frame field with respect to the natural one, although, of course, there are motions involved than just the rotations, since $\bar{v}^\mu_{,\nu} \in \mathfrak{gl}(n;\mathbb{R})$.

Equations (8.10) shows us that, in effect, there is more information in $D\Phi^\mu$ than there is in $D\mathbf{v}$, which should be obvious from the fact that $D\Phi^\mu$ defines a frame field, but $D\mathbf{v}$ does not have to have maximal rank as a matrix; indeed, it might vanish.

From the vanishing of $c^0_{ij}$, we can see that a frame field such as $\mathbf{e}_\mu$ would be unsuitable for representing a non-integrable complementary sub-bundle $\Sigma(M)$ to $L(M)$ since the sub-bundle that is spanned by $\{\mathbf{e}_i, i = 1, \ldots, n\}$ is clearly integrable.

Therefore, suppose that we have complementary sub-bundle $\Sigma(M)$ to $L(M)$ that is not integrable, and that its fibres are all annihilated by a non-zero 1-form $\phi$. Since it is complementary to $\mathbf{v}$ we also have that $\phi(\mathbf{v}) = \phi(\mathbf{e}_0) = \phi_0 \neq 0$. We can express $\phi$ with respect to the reciprocal coframe field $\theta^\mu$ to the $\mathbf{e}_i$ frame field as $\phi = \phi_\mu \theta^\mu$. The coframe members can be further related to the natural coframe field $dx^\mu$ by way of:

$$\theta^0 = dx^0, \qquad \theta^i = \tilde{\Phi}^i_{,j} dx^j. \tag{8.12}$$

When we project the $n$-frame field $\mathbf{e}_i$ onto the fibres of $\Sigma(M)$ we obtain a new $(n+1)$-frame field $\bar{\mathbf{e}}_\mu$:

$$\bar{\mathbf{e}}_0 = \mathbf{e}_0 = \mathbf{v}, \qquad \bar{\mathbf{e}}_i = -\phi_i \mathbf{e}_0 + \mathbf{e}_i. \tag{8.13}$$

We can obtain the non-zero structure functions for the $\bar{\mathbf{e}}_\mu$ frame field by direct computation from (8.13):

$$\bar{c}^0_{0i} = -\mathbf{v}\phi_i - \bar{c}^j_{0i}\phi_j, \qquad \bar{c}^j_{0i} = c^j_{0i} = \Phi^k_{,i,0}\tilde{\Phi}^j_{,k}, \tag{8.14a}$$

$$\bar{c}^0_{ij} = \phi_{[i}\mathbf{v}\phi_{j]} - \Phi^k_{,[i}\phi_{j],k} + \bar{c}^k_{ij}\phi_k, \qquad \bar{c}^k_{ij} = c^k_{ij} = \Phi^l_{,[i}\Phi^m_{,l],j}\Phi^k_{,m}. \tag{8.14b}$$

From the non-vanishing of $\bar{c}^0_{ij}$, the $\bar{\mathbf{e}}_\mu$ frame field is semi-holonomic.

We then achieve our objective of describing the transversal geometry of the flow in terms of the flow itself. We can obtain the torsion and curvature of the normal bundle by inserting the explicit expressions for the structure constants into the structure equations:

$$\Theta^k = -\tfrac{1}{2}(\Phi^l_{,[i}\Phi^m_{,l],j}\Phi^k_{,m})\,\theta^i \wedge \theta^j, \tag{8.15a}$$

$$\Omega = -\tfrac{1}{2}(\phi_{[i}\mathbf{v}\phi_{j]} - \Phi^k_{,[i}\phi_{j],k})\,\theta^i \wedge \theta^j + \Theta^i\phi_i. \tag{8.15b}$$

Since the line field $L(M)$ has rank one, the co-torsion and co-curvature vanish identically.

The structure functions $\bar{c}^0_{0i}$ and $\bar{c}^j_{0i}$ express how the fibres of the transverse bundle vary along the flow of $\mathbf{v}$. Up to sign, the former functions define a sort of covariant derivative of the 1-form $\phi$ with respect to the vector field $\mathbf{v}$:

$$\hat{\nabla}_{\mathbf{v}}\phi = (\mathbf{v}\phi_i + \bar{c}^j_{0i}\phi_j)\,\theta^i, \tag{8.16}$$

whereas the latter ones give the Bott connection of transverse vector fields with respect to $\mathbf{v}$:

$$\overset{\circ}{\nabla}_{\mathbf{v}}\mathbf{e}_i = \Phi^k_{,i,0}\tilde{\Phi}^j_{,k}\,\mathbf{e}_j. \tag{8.17}$$

We can also express the $\bar{\mathbf{e}}_\mu$ frame field in terms of the natural frame field for the adapted coordinates:

$$\bar{\mathbf{e}}_0 = \partial_0, \qquad \bar{\mathbf{e}}_i = -\phi_i\,\partial_0 + \Phi^j_{,i}\partial_j, \tag{8.18}$$

which puts the transition matrix $h^\mu_\nu$ for this transformation into the form:

$$h^\mu_\nu = \begin{bmatrix} 1 & -\phi_i \\ 0 & \Phi^i_{,j} \end{bmatrix}. \tag{8.19}$$

In this form, we can see that what the process of adapting the frame to $\Sigma(M)$ has done is to replace the zeroth component $w^0 = 1$ of any tangent vector field $\mathbf{w}$ with $\bar{w}^0 = 1 - \phi_i w^i$.

Of course, in the eyes of relativity the foregoing discussion was all non-relativistic, due to its reliance upon the proper time parameter. However, this is typical of working in the rest space of an observer.

Now, suppose our manifold $M$ has a Lorentzian structure $g$ for which the vector field $\mathbf{v}$ is timelike and proper time parameterized – so $g(\mathbf{v}, \mathbf{v}) = -1$, – and the sub-bundle $\Sigma(M)$ is the orthogonal complement to the line bundle generated by $\mathbf{v}$. If we assume that the adapted semi-holonomic frame field $\bar{\mathbf{e}}_\mu$ is orthonormal – so $g(\bar{\mathbf{e}}_\mu, \bar{\mathbf{e}}_\nu) = \eta_{\mu\nu} = \text{diag}(+1, -1,$

−1, −1) – then by direct calculation, we find that the components of $g$ with respect to the adapted natural frame field $\partial_\mu$ are:

$$g_{00} = 1, \qquad g_{0i} = \phi_i, \qquad g_{ij} = \phi_i \phi_j - \tilde{\Phi}^k_{,i} \tilde{\Phi}^k_{,j}. \tag{8.17}$$

## 9. 1+3 splittings and nonlinear connections.

One can specialize the methodology of sections 5 and 8 to the case in which $M$ is four-dimensional and the almost-product structure is defined by a 1+3 splitting of $T(M)$ into $L(M) \oplus \Sigma(M)$, with the rank-one temporal bundle $L(M)$ being orientable and generated by a global non-zero vector field **t**, which represents some physical motion.

If one also wishes to assume that $M$ is a Lorentzian manifold with a pseudo-metric $g$ on $T(M)$ then one might also wish to impose the constraints that **t** be timelike:

$$t^2 \equiv g(\mathbf{t}, \mathbf{t}) = 1, \tag{9.1}$$

and that the bundle $\Sigma(M)$ is orthogonal to $L(M)$, hence, spacelike. Consequently, one would then be assuming that:

$$0 = g(\mathbf{v}, \mathbf{t}) \qquad \text{for all } \mathbf{v} \in \Sigma(M). \tag{9.2}$$

Since $L(M)$ is integrable, we regard it as representing a generalization of the vertical bundle on $GL(M)$ and the complementary sub-bundle $\Sigma(M)$ will then represent the analogue of a horizontal sub-bundle on $GL(M)$ that one defines by a choice of linear connection [4]. Hence, we shall not necessarily assume that $\Sigma(M)$ is integrable. It is then best to think of the temporal line field $L(M)$ as the structural sub-bundle and the spatial complement $\Sigma(M)$ as a representation of its normal bundle as a transversal sub-bundle in $T(M)$.

If we assume that there is a global flow for the line field $L(M)$ – viz., a global action of $(\mathbb{R}, +)$ on $M$ – then we can think of the resulting foliation by integral curves as a generalization of a $G$-principal bundle in which we have sacrificed the existence of a fibration, which would imply the existence of a global slice for the flow, together with the assumption that all of its orbits were diffeomorphic, or the assumption that the quotient space $M/\mathbb{R}$ – i.e., the orbit space of the group action or the leaf space of the foliation – admits the structure of a differentiable manifold. The action of $\mathbb{R}$ is by proper time translation along the integral curves of the foliation and the fundamental vector field of this action is then **t** itself.

We shall then think of a semi-holonomic almost-product structure on $M$ that admits a group action on the leaves of its foliation as a *generalized gauge structure* for the motion.

It is essential for the physical interpretation to discuss the physical nature of the non-integrability assumption on $\Sigma(M)$. If $\Sigma(M)$ were integrable as a differential system then

---

[4] Since we are saying that $\Sigma(M)$ is analogous to a vertical, or structural, sub-bundle and $L(M)$, to a horizontal, or transversal, sub-bundle, it would seem more consistent to refer to a 3+1 splitting $\Sigma(M) \oplus L(M)$ of $T(M)$; nevertheless, we hope that no confusion will arise.

*M* would admit a codimension-one foliation by leaves that would represent the proper-time simultaneous points in spacetime relative to the motion defined by **t**. Hence, we are considering the possibility that the motion defined by **t** admits only a singular proper time foliation, such as one might encounter with vortical motions or cosmological models in which the singularities of spacetime also obstruct the integrability of the Cauchy problem for the proper time evolution of the initial spatial metric. We should point out that since most cosmological models assume from the outset that this initial-value problem in geometrodynamics is integrable, so spacetime becomes the four-dimensional evolute of a three-dimensional initial space $\Sigma$, it is only to be expected that the only foliation that cosmology generally considers is the "cylindrical" foliation that takes the form of $\mathbb{R} \times \Sigma$.

In particular, a maximal Cauchy hypersurface is a global slice for the proper time flow that carries the initial data into the future.

In addition to assuming that $L(M)$ is generated by a global non-zero vector field **t**, we also assume that $\Sigma(M)$ is annihilated by a global non-zero 1-form $\theta$ with $\theta(\mathbf{t}) \neq 0$; i.e., it is the (algebraic) solution to the exterior differential equation:

$$\theta = 0. \tag{9.3}$$

The assumption of non-integrability for $\Sigma(M)$ implies that:

$$\Omega \equiv d\theta = \eta \wedge \theta + \sigma \tag{9.4}$$

for some $\eta \in \Lambda^1(M)$ and some non-zero $\sigma \in \Lambda^1(\Sigma)$.

When $\theta$ is the *covelocity* 1-form associated with the velocity vector field **t** by way of *g*:

$$\theta = i_\mathbf{t} g = (g_{\mu\nu} t^\nu)\theta^\mu, \tag{9.5}$$

then $\Omega$ is the kinematical vorticity of the flow of **t**. (In (9.5), we have abbreviated the usual introduction of a local coframe field $\theta^\mu$ on an open subset $U \subset M$ down to an equal sign.) In the more general, non-metric, case, we shall think of $\theta$ as a nonlinear connection 1-form with values in $\mathbb{R}$ and $\Omega$ as its curvature 2-form. When *g* is expressed in a frame field that is adapted to **t** – so $\mathbf{e}_0 = \mathbf{t}$ and $g_{\mu 0} = \delta_{\mu 0}$ – one has that $\theta = \theta^0$.

The Frobenius 3-form, which dictates the integrability of $\Sigma(M)$, is simply:

$$\mathcal{I} = \theta \wedge d\theta = \theta \wedge \Omega; \tag{9.6}$$

it is related to the Pauli-Lubanski spin vector, which starts with the energy-momentum 1-form instead of covelocity.

Between **t** and $\theta$, we can construct the temporal projector field *L*, the spatial projector field $\Sigma$, and the almost-product field *F* by way of:

$$L = \theta \otimes \mathbf{t}, \qquad \Sigma = I - \theta \otimes \mathbf{t}, \qquad F = I - 2\theta \otimes \mathbf{t}. \tag{9.7}$$

This means that $L$ is the connection 1-form for the splitting in the sense of Fröhlicher-Nijenhuis. It is sometimes [3, 4, 19] referred to as an *observer field.*

The associated Fröhlicher-Nijenhuis torsion, curvature, co-torsion, and co-curvature 2-forms then take the form:

$$\Theta = -\Sigma[\Sigma, \Sigma], \qquad \Omega = -L[\Sigma, \Sigma], \qquad (9.8a)$$

$$\tilde{\Theta} = -L[L, L] = -[L, L], \qquad \tilde{\Omega} = -\Sigma[L, L] = 0, \qquad (9.8b)$$

which gives:

$$\Theta(\mathbf{v}, \mathbf{w}) = -\Sigma[\mathbf{v}_\Sigma, \mathbf{w}_\Sigma], \qquad \Omega(\mathbf{v}, \mathbf{w}) = -\theta[\mathbf{v}_\Sigma, \mathbf{w}_\Sigma], \qquad (9.9a)$$

$$\tilde{\Theta}(\mathbf{v}, \mathbf{w}) = -[\mathbf{v}_L, \mathbf{w}_L], \qquad \tilde{\Omega}(\mathbf{v}.\mathbf{w}) = -\Sigma[\mathbf{v}_L, \mathbf{w}_L] = 0 \qquad (9.9b)$$

for arbitrary vector fields $\mathbf{v}, \mathbf{w} \in \mathfrak{X}(M)$.

Now, let $\{\mathbf{e}_\mu, \mu = 0, \ldots, 3\}$ be an adapted local frame field on $U \subset M$ and let $\{\theta^\mu, \mu = 0, \ldots, 3\}$ be its reciprocal coframe field. Since it is adapted, there is no loss in generality in assuming that $\mathbf{e}_0 = \mathbf{t}$ and $\theta^0 = \theta$. We can then give local expressions for (9.9a, b):

$$\Theta^i = d\theta^i = -\tfrac{1}{2} c^k_{ij}\, \theta^j \wedge \theta^k, \qquad \Omega = d\theta = -\tfrac{1}{2} c^a_{ij}\, \theta^i \wedge \theta^j, \qquad (9.10a)$$

$$\tilde{\Theta} = 0, \qquad \tilde{\Omega}^i = -\tfrac{1}{2} c^i_{00}\, \theta \wedge \theta = 0. \qquad (9.10b)$$

However, as discussed above, $\mathbf{e}_\mu$ cannot be a natural frame field for any coordinate system $(U, x^0, x^i)$, so the closest one can come to an adapted coordinate system is a semi-holonomic one, for which $x^0$ is the adapted coordinate; i.e., $\mathbf{e}_0 = \partial_0$. Hence, one must express $\mathbf{e}_\mu$ with respect to $\partial_0$ by means of:

$$\mathbf{e}_0 = \partial_0, \qquad \mathbf{e}_i = \partial_i - \theta_i\, \partial_0. \qquad (9.11)$$

The reciprocal semi-holonomic coframe field then takes the form:

$$\theta^0 = \theta = dx^0 + \theta_i\, dx^i, \qquad \theta^i = dx^i. \qquad (9.12)$$

In order to obtain the structure functions of a semi-holonomic adapted local frame field $\mathbf{e}_\mu$, we specialize the general result (3.2) to obtain the non-vanishing commutation relations in the form:

$$[\mathbf{e}_i, \mathbf{e}_j] = -(\mathbf{e}_i \theta_j - \mathbf{e}_j \theta_i)\, \mathbf{e}_0, \qquad (9.13a)$$

$$[\mathbf{e}_0, \mathbf{e}_i] = -(\mathbf{e}_0 \theta_i)\, \mathbf{e}_0. \qquad (9.13b)$$

We then deduce that in a semi-holonomic frame the structure equations take the form of (9.10a, b), in which the non-zero structure functions are:

$$c^0_{ij} = -(\mathbf{e}_i \theta_j - \mathbf{e}_j \theta_i), \qquad c^0_{0i} = -\mathbf{e}_0 \theta_i. \qquad (9.14)$$

We obtain the co-torsion 2-form from the fact that $c^0_{00}$ must vanish, and equation (9.13b) gives us the vanishing of the co-curvature 2-form:

$$\tilde{\Theta} = 0, \qquad \tilde{\Omega}^i = 0. \qquad (9.15)$$

We can interpret the relationship that is expressed by (9.13b) in various ways: If one thinks of the spatial 3-frame $\mathbf{e}_i$ as being convected along the flow of $\mathbf{e}_0$ then it says:

$$L_{\mathbf{e}_0} \mathbf{e}_i = \lambda_i \mathbf{e}_0, \qquad (\lambda_i \equiv -\mathbf{e}_0 \theta_i), \qquad (9.16)$$

i.e., that the proper time derivatives of the 3-frame members are all in the proper time direction. On the other hand, if one thinks of the vector $\mathbf{e}_0$ as being convected along the flows of the vector fields $\mathbf{e}_i$ then one has:

$$L_{\mathbf{e}_i} \mathbf{e}_0 = -\lambda_i \mathbf{e}_0, \qquad (9.17)$$

which says that $\mathbf{e}_0$ is an eigenvector of the Lie derivative operator in each of the directions of the three-frame.

Furthermore, since we are also requiring that the transversal part of $L_{\mathbf{e}_0} \mathbf{e}_i$ vanish in (9.13b), we can express this condition in terms of the covariant derivative that is defined by the Bott connection for the foliation defined by $L(M)$:

$$0 = \overset{o}{\nabla}_{\mathbf{e}_0} \mathbf{e}_i = H[\mathbf{e}_0, \mathbf{e}_i]. \qquad (9.18)$$

Hence, a semi-holonomic local frame field will be parallel along the flow of $\mathbf{t}$ with respect to the Bott connection; i.e., it is *Bott parallel.*

If we define a local frame field $\overline{\mathbf{e}}_\mu$ on $U$ that is convected along the flow of $\mathbf{t}$ and adapted to a non-integrable $\Sigma(M)$ then we can read off the necessary information for describing the geometry of the sub-bundle $\Sigma(U)$, as seen be an observer moving with $L(U)$, from the expressions in (8.14a,b). We adjust the notation accordingly by replacing $\mathbf{v}$ with $\mathbf{t}$ and $\phi$ with $\theta$:

$$\overline{c}^0_{0i} = -\mathbf{t}\phi_i - \overline{c}^j_{0i}\theta_j, \qquad \overline{c}^j_{0i} = c^j_{0i} = \Phi^k_{,i,0}\tilde{\Phi}^j_{,k}, \qquad (9.19a)$$

$$\overline{c}^0_{ij} = \theta_{[i}\mathbf{t}\theta_{j]} - \Phi^k_{,[i}\theta_{j],k} + \overline{c}^k_{ij}\theta_k, \qquad \overline{c}^k_{ij} = c^k_{ij} = \Phi^l_{,[i}\Phi^m_{,l],j}\Phi^k_{,m}. \qquad (9.19b)$$

The second equation in (9.19a) tells us about the Bott connection on $\Sigma(U)$. A special role is then attached to pairs $\{\mathbf{t}, \theta\}$ for which $\overline{c}^0_{0i}$ vanishes, since that becomes the condition:

$$\mathbf{t}\theta_i + \overline{c}^j_{0i}\theta_j = 0, \qquad (9.19)$$

which amounts to the vanishing of a covariant derivative of the spatial 1-form described by $\theta_i$ if one uses the Bott connection, whose components in the $\overline{\mathbf{e}}_\mu$ frame field are given by $\overline{c}^j_{0i}$. One could then call such a pair $\{\mathbf{t}, \theta\}$ Bott parallel, or, for that matter, the observer field $\theta \otimes \mathbf{t}$ that they define.

Equations (8.15a, b) give us the curvature and torsion of the nonlinear connection on the bundle $\Sigma(U)$ that is defined by the observer field $\theta \otimes \mathbf{t}$ as they appear in an adapted convected frame field:

$$\Theta^k = -\tfrac{1}{2}(\Phi^l_{,[i}\Phi^m_{,l],j}\Phi^k_{,m})\,\theta^i \wedge \theta^j, \tag{9.20a}$$

$$\Omega = -\tfrac{1}{2}(\theta_{[i}\,\mathbf{t}\theta_{j]} - \Phi^k_{,[i}\theta_{j],k})\,\theta^i \wedge \theta^j + \Theta^i\theta_i. \tag{9.20b}$$

We can specialize the discussion of horizontal curves in section 6.4 to a 1+3 splitting in a straightforward way. In the present case, since the vertical bundle has one-dimensional fibres a curve $\sigma: (-\varepsilon, +\varepsilon) \to M$ is horizontal iff $\dot{\sigma}(\tau) \in \Sigma(M)$ for all $\tau$. This condition then takes the local form:

$$\theta(\dot{\sigma}(\tau)) = 0 \qquad \text{for all } \tau. \tag{9.21}$$

Let $(x^0, x^i)$ be a coordinate system on $U$ that is adapted to $L(U)$ and let $\{\partial_0, \partial_i\}$ be its natural frame field. Now, suppose that $\{\mathbf{e}_0, \mathbf{e}_i\}$ is an adapted semi-holonomic local frame field on $U$ that is related to $\{\partial_0, \partial_i\}$ by way of (9.11).
If:

$$\dot{\gamma} = v^i\,\mathbf{e}_i = \dot{x}^0\partial_0 + \dot{x}^i\partial_i \tag{9.22}$$

then (9.20) gives rise to the following system of ordinary differential equations for $\sigma$:

$$\dot{x}^0 - \theta_i\dot{x}^i = 0. \tag{9.23}$$

The best way to apply the concept of horizontal curve in this sense to physical motions is to think of a horizontal vector field $\mathbf{v}(\tau)$ along a curve $\chi(\tau)$ in a congruence that describes some physical motion as representing a transversal variation $\delta\chi(\tau)$ of the curve to an infinitesimally close curve that takes each point of the initial curve to a proper-time simultaneous point on the neighboring curve. If the congruence represents curves of least action, in some sense, then one can then regard the constraint $\theta(\delta\gamma) = 0$ as something to be included in the formulation of the second variation, such as the problem of geodesic deviation.

This shows how one can interpret the torsion and curvature of the nonlinear connection whose 1-form is $\theta \otimes \mathbf{t}$: Suppose one has two horizontal vector fields $\mathbf{X}$ and $\mathbf{Y}$ that defined on a neighborhood of some point $x \in M$. If one follows the integral curves that start at $x$ and proceed around an infinitesimal parallelogram spanned by $\mathbf{X}$ and $\mathbf{Y}$ then the value of the torsion $\Theta^i(\mathbf{X}, \mathbf{Y})$ shows how much the endpoint of the curve will be infinitesimally translated in the transverse direction and the value of the curvature $\Omega(\mathbf{X}, \mathbf{Y})$ says how much it will be infinitesimally translated in the direction of the integral curve through $x$. We illustrate this in figure 1.

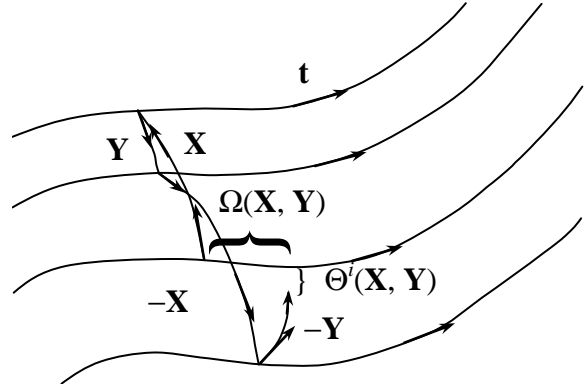

Figure 1. Effect of torsion and curvature on an infinitesimal horizontal parallelogram.

## 10. Discussion.

One obvious direction to proceed in the study of nonlinear connections on spacetime is perhaps the conventional one of introducing a Lorentzian structure and its Levi-Civita connection in such a way that both are adapted to the existence of a 1+3 almost-product splitting of spacetime. One can then examine the way that the geometry of the manifold, as described by the Levi-Civita connection, its curvature, and its geodesics, appear in the spatial and temporal sub-bundles. In a subsequent study, we shall define this problem in the language of partial connections on foliated bundles. In particular, the bundle is the bundle $SO(3,1)(M)$ of Lorentzian frames and the foliation is defined by the lift of the foliation of **t** on $M$ to a foliation of $SO(3,1)(M)$. A particular case of a foliation of $M$ that then becomes most interesting is that of a geodesic congruence relative to the Levi-Civita connection.

However, there are many more topics that relate to nonlinear connections that one can investigate without introducing another connection. For instance, before one considers geodesic congruences one might consider a generalization of the process of second variation to transformations of a foliation that preserve the leaves. In the case of a congruence this suggests transverse vector fields that generate transformations that take curves of the congruence to neighboring curves of the congruence. It is possible that some generalization of the equations of geodesic deviation might exist that involves the curvature of the nonlinear connection that defines the flow.

There is also much to be investigated in the form of the transversal geometry of flows on a manifold when one defines a Riemannian structure on $\Sigma(M)$ – i.e., the normal bundle to the flow – but not necessarily a Lorentzian structure on $T(M)$ itself. This might provide more insight into the process of extending the Euclidian geometry of one's intuition to the four-dimensional geometry of spacetime than one usually deduces from conventional special relativity. In particular, as discussed in [**20**], this process has considerable projective geometric significance that is not usually emphasized.

Furthermore, there is much to recommend the study of singular foliations that are defined by flows with fixed points – i.e., the vector field **t** would have zeroes. The information that is contained in the stability matrix at a fixed point – which is defined by $D\mathbf{t}$ – will affect the geometric information that is transversal to the flow, since it relates to the second derivatives of the flow at a fixed point. One expects that such fixed points are inevitable if singularities also entail a breakdown of time-orientability at such points.